\documentstyle[12pt,epsf]{article}
\def\lsim{\mathrel{\rlap{\lower3pt\hbox{\hskip0pt$\sim$}}
    \raise1pt\hbox{$<$}}}         
\def\gsim{\mathrel{\rlap{\lower4pt\hbox{\hskip1pt$\sim$}}
    \raise1pt\hbox{$>$}}}         
\begin{document}
\renewcommand{\theequation}{\thesection.\arabic{equation}}
\newcommand{\beq}{\begin{equation}}
\newcommand{\eeq}{\end{equation}}
\def\beqn{\begin{eqnarray}}
\def\eeqn{\end{eqnarray}}
\newcommand{\qt}{\tilde q}
\newcommand{\Tr}{{\rm Tr}\,}
\newcommand{\E}{{\cal E}}
\newcommand{\qtu}{\tilde q_{1}}
\newcommand{\qtd}{\tilde q_{2}}
\newcommand{\ntwo}{${\cal N}=2\;$}
\newcommand{\none}{${\cal N}=1\;$}
\newcommand{\ax}{\vert\xi\vert}
\newcommand{\xva}{\vert\vec{\xi}\vert}
\newcommand{\rt}{\tilde r}
\newcommand{\ar}{i\theta}
\newcommand{\as}{\alpha^{2}}
\newcommand{\gs}{g^{2}}
\newcommand{\fl}{\phi_{l}}
\newcommand{\fh}{\phi_{h}}
\newcommand{\pu}{\psi_{1}}
\newcommand{\pd}{\psi_{2}}
\newcommand{\vp}{\varphi}
\newcommand{\ve}{\varepsilon}
\newcommand{\pt}{\partial}
\newcommand{\pz}{\partial_{z}}

\begin{titlepage}
\renewcommand{\thefootnote}{\fnsymbol{footnote}}

\begin{flushright}
TPI-MINN-02/50\,,\,
UMN-TH-2124/02\\
ITEP-TH-77/02

\end{flushright}

\vfil

\begin{center}
\baselineskip20pt
{ \Large \bf Domain Walls and Flux Tubes
   in \boldmath{\ntwo} SQCD: D-Brane Prototypes}
\end{center}
\vfil

\begin{center}

\vspace{0.3cm}

{\large

{ \bf    M.~Shifman$^{a}$} and { \bf A.~Yung$^{a,b,c}$}
}

\vspace{0.3cm}

$^a${\it William I. Fine Theoretical Physics Institute, University of Minnesota,
Minneapolis, MN 55455, USA}\\
$^b${\it Petersburg Nuclear Physics Institute, Gatchina, St. Petersburg
188300, Russia}\\
$^c${\it Institute of   Theoretical and Experimental Physics,
Moscow  117250, Russia}\\

\vfil

{\large\bf Abstract} 
\vspace*{.25cm}

\end{center}

This paper could have been entitled ``D branes and strings from flesh and blood."
We study field theoretic prototypes of D branes/strings. To this end
 we consider (2+1)-dimensional domain walls
in  (3+1)-dimensional \ntwo SQCD with  SU(2)  gauge group and $N_f=2$ flavors of
fundamental hypermultiplets (quarks). This theory is perturbed by a small mass
term of the adjoint matter which, in the leading order in the mass parameter,
does not break \ntwo supersymmetry, and reduces to a (generalized) Fayet-Iliopoulos term
in the effective low-energy \ntwo SQED. We find   1/2 BPS-saturated domain wall solution 
interpolating between two quark vacua at  weak coupling, and show that this
domain wall localizes a U(1)  gauge field. To make contact with the 
brane/string picture we consider the Abrikosov-Nielsen-Olesen magnetic flux
tube in one of two quark vacua and demonstrate that it can end on the domain wall.
We find an explicit 1/4 BPS-saturated solution for the wall/flux tube junction.
We verify that the end point of the flux tube on the wall  plays the role of an electric
charge in the dual  2+1  dimensional SQED living on the wall. Flow to   \none
theory is   discussed. Our results lead us to a conjecture
regarding the notorious
``missing wall" in the solution of Kaplunovsky {\em et al.}

\vfil

\end{titlepage}

\newpage

\section{Introduction}
\label{introduction}

D branes are extended objects in string theory on which strings can end \cite{P}.
Moreover, the gauge fields are the lowest excitations 
of open superstrings, with the end points attached to  D branes.
 SU$(N)$ gauge theories are obtained as a field-theoretic reduction
of a string theory  on the world volume of a stack of $N$ D branes.

In the recent years solitonic objects of the domain wall and string type
were extensively studied  in supersymmetric gauge theories in 1+3 dimensions.
First, it was observed \cite{DS} that there should
exist critical (BPS saturated) domain walls
in \none  gluodynamics, with the tension scaling as
$N\Lambda^3$. (Here   $\Lambda$ is the scale parameter.) 
The peculiar $N$ dependence of the tension 
 prompted \cite{Witten:1997ep}
a D brane interpretation of such walls. Ideas as to how
flux tubes can end on the BPS walls were analyzed \cite{Kogan:1997dt}
at the qualitative level shortly thereafter. Later 
on, BPS saturated strings and their junctions with domain walls
were discussed 
\cite{HSZ,VY,GPTT} in a more quantitative aspect in \ntwo theories.
Some remarkable parallels between field-theoretical critical solitons
and the D-brane construction were discovered. In this paper we undertake a systematic
investigation of this issue --- parallel between  field-theoretical critical solitons
and D branes/strings. The set up which will  provide  us with
multiple tools useful in this endeavor is  \ntwo SQCD 
considered by Seiberg and Witten \cite{SW1,SW2}. Following the original publications,
we will introduce a  parameter $\mu$ which explicitly breaks
\ntwo down to ${\cal N}=1$. It turns out that in the limit $\mu\ll\Lambda$
we will be able to verify many of the previous conjectures as well as establish new results
in the reliable regime of weak coupling.

Research on field-theoretic mechanisms of gauge field localization
on the domain walls is an important ingredient of our analysis.
The only viable mechanism of gauge field localization
was put forward in Ref. \cite{DS}
where it was noted that if a gauge field is confined in the bulk and
is unconfined (or less confined) on the brane, this naturally gives rise
to a gauge field on the wall (for further developments see Refs. \cite{DR,DV}).
Although this idea seems easy to implement, in fact
it requires a careful consideration of quantum effects
(confinement is certainly such an effect)
which is hard to do  at strong coupling.
This again leads us to models in which the gauge field 
localization can be implemented at weak coupling. 
We use, in addition,  some new general results \cite{AVafa}
(see also Ref. \cite{Ritz:2002fm})
regarding effective field theories on the critical domain walls
in supersymmetric gluodynamics. In the present paper we focus
on localization of the Abelian gauge field.
The issue of  non-Abelian gauge fields on domain walls
will be addressed in  the subsequent publication.

Our main results can be summarized as follows. 
First, we suggest an \ntwo model 
(\ntwo may or may not be softly broken down
to \none) which possesses
both critical walls and strings, at weak coupling.
In this model one can address all questions 
regarding the gauge field localization on the wall
and the wall-string junction,
 and answer these questions in a fully controllable
manner. We find that there exists an 1/2 BPS domain wall which does
localize a U(1) gauge field; the charge which presents
 the source for this
field is confined in the bulk. We find that an 1/2 BPS
flux tube coming from
infinity does indeed end on the above wall.
The wall-string junction is 1/4 BPS. When the string ends on the wall
the latter is no more flat, it acquires a logarithmic bending
which is fully calculable.

In more detail, our theoretical set-up can be described  as follows.
 We consider  (2+1)-dimensional critical domain walls in 
(3+1)-dimensional SU(2)   SQCD 
originally studied by Seiberg and Witten \cite{SW1,SW2}. 
To ensure the existence of 
  domain walls at weak coupling we introduce 
$N_f=2$ flavors of fundamental  (quark) hypermultiplets. 
This theory has a Coulomb
branch on which the adjoint scalar acquires an arbitrary 
vacuum expectation value (VEV), 
$$
\langle \Phi\rangle = \langle a \rangle \, \frac{\tau_3}{2}\,, 
$$
 breaking the
 SU(2)  gauge group down to  U(1). The Coulomb branch has four singular points
in which either monopole, or dyon or one of the two quarks become massless. 
The first two of these points
are always at strong coupling,  while the massless quark points can be at weak coupling provided
that
the quark mass parameters $m_A$ are large, $m_A\gg \Lambda$, where $A=1,2$ is the flavor index.
Below the vacua in which quarks become massless will be referred
to as the quark vacua.

In order to have domain walls,
the vacuum manifold, rather than
being continuous,  must consist of isolated points. To guarantee the existence of
discrete vacua
 we perturb the above theory by adding a small
mass term for the adjoint matter,  via superpotential
\beq
\label{brsup}
{\cal W}_{u}=\mu\, u \,,\qquad  u \equiv \Tr \Phi^2\,.
\eeq
 
Generally speaking, the superpotential breaks
\ntwo down to ${\cal N}=1$.
The Coulomb branch shrinks to
four above-mentioned isolated \none vacua.
Of  special  importance for what follows is the fact that
 \ntwo supersymmetry is {\em not}
broken \cite{HSZ,VY,Hou} to the leading order in the parameter $\mu$
in the effective theory. 
In the effective low-energy SQED the superpotential
(\ref{brsup}) gives rise to a superpotential linear in $a$ plus
 higher order corrections.
If only the linear term  in $a$ in the superpotential
is kept, the theory is exactly \ntwo.

We will be mostly interested in the quark vacua
since they yield weak coupling regime. Near the
quark vacua, to the leading order in \ntwo breaking parameter,
the superpotential in the effective low-energy SQED is
\beq
\label{fi}
W_{\rm SQED} = -\frac1{2\sqrt{2}}\,\xi\, a ,
\eeq
where the coefficient $\xi$ is determined by the VEV of the lowest component of $a$
 in the given quark vacuum,
\beq
\label{xi}
\xi=-2\sqrt{2}\mu \langle a\rangle\,,\qquad \langle a\rangle\sim m_A\gg\mu\,.
\eeq
The perturbation (\ref{fi}) can be ``rotated"  \cite{VY} in such a way  as 
to render it  a Fayet-Iliopoulos (FI) term \cite{FI};
  {\em per se} it does not break  \ntwo supersymmetry.
If $\xi\neq 0$,  the quark fields develop VEV's (of order of $\sqrt\xi$) breaking
 U(1)  gauge symmetry, so that the theory becomes fully higgsed.
Then  we consider a domain wall interpolating between the two quark vacua,
a task which can be addressed  at weak coupling. We 
also analyze the string-wall junction.

Our domain wall is  1/2 BPS-saturated.
 It turns out that the solution of the first-order Bogomolny equations
can
be readily found in the range of parameters
\beq
\label{mxi}
\sqrt{\xi}\ll \Delta m \ll m_1 \sim m_2\, ,\qquad  \Delta m\equiv m_1-m_2\,,
\eeq
when two quark vacua come close   to each other on the would-be 
Coulomb branch.
Qualitatively the solution has the following structure: the quark fields are
small inside the wall, while $a$ is a slowly varying (almost linear) function of $z$
where $z=x_3$ is the coordinate orthogonal to the wall.
The original U(1) gauge field is higgsed outside the wall --- this
 is a ``superconducting" phase.
Inside the wall superconductivity is destroyed.
Correspondingly, magnetic charges are confined in the bulk \cite{HSZ,Ymc4},
giving rise to magnetic flux tubes in the bulk\,\footnote{
The magnetic flux tube is nothing but
the Abrikosov-Nielsen-Olesen (ANO) string \cite{ANO}. Flux tubes in the Seiberg-Witten
theory were studied in \cite{FG,Ymc4,EFMG,VY,KoS}.}, while inside the wall
the magnetic flux can spread freely.
The  U(1) gauge field $A_m^{(2+1)}$ localized on the wall,  which describes
interaction of the probe magnetic charges placed on the wall, is dual to the
original U(1) gauge field.

As well-known \cite{Polyakov:1976fu}, 2+1 dimensional gauge field is equivalent to
a real scalar (compact) field --- we will call it $\sigma$. This must
 be one of the moduli fields.
In the limit of \ntwo supersymmetry, the effective field theory
on the 1/2 BPS wall must possess four conserved supercharges
(i.e. it is \ntwo from the (2+1)-dimensional standpoint).
Then the minimal supersymmetry representation contains two
real boson fields ---
the effective field theory
on the   wall must include two
real boson fields.
One is the above mentioned  $\sigma (t,x,y)$, another originates
from the translational collective coordinate, the
 position of the
wall $z_0$. We will refer to this field as $\zeta  (t,x,y)$. 

In the limit of exact \ntwo, the field $\sigma (t,x,y)$ is massless,
 as well as $\zeta  (t,x,y)$,
and is related to the
gauge field strength tensor as follows:
\beq
\label{21gauge}
F^{(2+1)}_{nm}={\rm const}\cdot \,\varepsilon_{nmk}\, \partial^k \sigma\,,
\eeq
where $n,m=0,1,2$ and the constant on the right-hand side has dimension of mass.
Taking account of higher orders in $\mu$ 
(i.e. quadratic in $a-\langle a\rangle$ term in the superpotential)
breaks \ntwo supersymmetry 
of our macroscopic theory down to \none\,. Surprisingly, 
this does not generate  a mass term for the field $\sigma (t,x,y)$,
which remains  a moduli field.
This can be seen in many different ways.
One of them is through analyzing   the fermion zero modes.
A Jackiw-Rebbi type index theorem
\cite{jackiw} 
tells us that fermion zero modes (those unrelated to the supertranslational ones)
exist even though \ntwo is broken. \none
supersymmetry  requires then a bosonic superpartner,
which is the massless field $\sigma (t,x,y)$. 
Thus, at the level of quadratic in derivative terms,
the effective moduli field theory on the
domain wall world volume is \ntwo (four supercharges).
The breaking presumably occurs if higher-derivative terms
are taken into account.

\begin{figure}[h]  
\epsfxsize=7cm
\centerline{\epsfbox{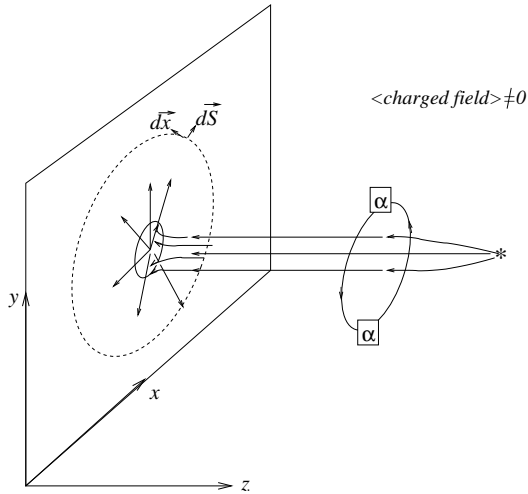}}
 \caption{Geometry of
the wall-string junction.
The gauge field localized on the wall
can be cast in the form of a dual 1+2 dimensional electrodynamics. This
field is dual   to the gauge field higgsed in the bulk.
The probe magnetic charge in the bulk 
(the magnetic monopole of the original \ntwo SQCD)
is denoted by asterisk.}
\label{syfigtwo} 
\end{figure}

Next, completing the theme of the gauge field localization on the wall we proceed
 to the second aspect of the problem --- the issue of how strings 
originating in the bulk can end on the wall.
In string theory, the brane localization of gauge fields   is closely related to
the possibility for an open  string to end  on a D brane. 
Since, as we assert, the BPS walls and flux tubes in \ntwo SQCD 
present a close prototype,  
it is instructive to study this phenomenon
in field theory. That the magnetic field flux tube
will end on the wall was already explained above, at a qualitative
level. 
There is no doubt that the phenomenon does take place in our model.
Our task is more quantitative, however.
We want to find (and do find) a 1/4 BPS solution of the first order
Bogomolny equations 
that describes an ANO flux tube ending on the wall. 
In other words, attaching a flux tube to 1/2 BPS wall makes the configuration
1/4 BPS.
The attachment of the tube gives rise to 
 two effects. First, the wall is now bent, and,   second, the field $\sigma (x,y)$
develops a vortex. If $\{x_0,y_0\}$ are the coordinates of the
tube center on the wall, 
at large separations $r$ from the center
\beq
\label{sigalph}
\sigma (x,y ) =\alpha\,, 
\eeq
where $\alpha$ is the polar angle on the two dimensional wall surface  
(Fig. \ref{syfigtwo}) and $ r=\sqrt{(x-x_0)^2 + (y-y_0)^2}$.

According to Eq. (\ref{21gauge}) the  vortex in $\sigma$
is equivalent to a radial   electric field in the dual 1+2 dimensional QED on the wall,
\beq
\label{elf}
F_{0i}^{(2+1)}={\rm const}\cdot \,\frac{x_i-(x_0)_i}{r^2}\,,\qquad i=1,2\,.
\eeq
 Thus, 
 the string end point on the wall plays the role of a probe charge for the 
dual 1+2 dimensional QED  on the wall.

Diminishing the mass parameter
$m^A$ we move the quark vacua toward the strong coupling
regime. (Simultaneously the parameter governing the breaking \ntwo $\to$
\none becomes larger.) At strong coupling there exist two additional vacua,
the monopole and dyon ones.
A domain wall interpolating between 
the monopole and dyon vacua  
was discussed previously \cite{KS}. This is a strong coupling
problem which requires a ``patchy" description ---
one cannot introduce one and the same
effective low-energy theory which would be  valid
simultaneously near the monopole and dyon vacua because monopoles and dyons
are not mutually local states.
Nevertheless, we have something new to say   regarding
the wall interpolating between 
the monopole and dyon vacua. The  
 analysis of Ref. \cite{KS} does not seem to be complete since it does
not take into account moduli dynamics on the wall.
As is well-established \cite{KSS,AVafa,Ritz:2002fm}, in SU(2) theory
two {\em distinct} critical domain walls (with one and the same
tension) must exist. At the same time, only one domain wall was detected
in Ref. \cite{KS}. We suggest a tentative
solution to  the paradox of a ``missing wall" which should emerge 
from consideration
of the moduli dynamics.

A
  problem similar to ours was addressed previously in Ref. \cite{GPTT}
in the context of
(3+1)-dimensional \ntwo massive sigma models
 on hyper-K\"{a}hler target spaces
 (see also Ref. \cite{GTT}).
The results we obtain in \ntwo SQCD 
are in qualitative agreement  with those  obtained in Ref.~\cite{GPTT}
in sigma models. 
In particular in sigma model   the  existence of
a compact moduli field on the wall,
representing dual to  U(1) gauge field, was demonstrated,
which is certainly
  no accident.
Indeed,   in the limit opposite to the that quoted in Eq.~(\ref{mxi}), when 
$\sqrt{\xi}\gg \Delta m $, the photon and its superpartners become heavy 
in \ntwo SQCD, and 
can be integrated out. Then, \ntwo SQCD reduces in the  low-energy limit
to \ntwo sigma model with the hyper-K\"{a}hler Eguchi-Hanson target space studied in
Ref.~\cite{GPTT}. 

The paper is organized as follows. In Sect. \ref{toymodel} we present 
a toy (nonsupersymmetric) model which has a domain wall and exhibits
the phenomenon of the gauge field 
localization on the wall.
Although this model is primitive it serves as a nice illustration
for one of the phenomena we are interested in ---
the occurrence of a (2+1)-dimensional gauge field localized on the
wall. Section \ref{basicmodel} introduces our basic model ---
supersymmetric QED --- obtained as a reduction of the Seiberg-Witten model. 
We specify in which limit  the model  has  extended \ntwo
supersymmetry, while for nonlimiting values of parameters
it is \none.
 In Sect. \ref{domainwall} we derive
and solve first order BPS equations for the domain wall interpolating between two
quark vacua. Of most importance is Subsec. \ref{kinterms}
where we derive field theory for the moduli fields
living on the wall. 
In Sect. \ref{anostrings} we review the ANO strings in (the low-energy limit of)
the Seiberg-Witten theory.
 Section \ref{stringend} treats the issue of strings ending on the wall.
Here we derive  first order BPS equations for the string-wall junctions
and 
discuss the properties
of  1/4 BPS solutions to these equations. 
We find how the magnetic flux which the flux tube
brings to the wall spreads out inside the wall.
In Sect. \ref{flow}  we discuss 
the impact of soft breaking of
\ntwo down to \none in our model.  Brief remarks on the literature,
including the mystery of a ``missing wall" in the solution of Kaplunovsky
{\em et al.} are presented in Sect. \ref{comments}.

\section{A toy model}
\label{toymodel}
\setcounter{equation}{0}

In this section we will consider a toy (non-supersymmetric)
model which exhibits the phenomenon we are interested in
--- the gauge field localization on a wall.
This model lacks certain ingredients
which will be of importance in the analysis
of more sophisticated supersymmetric models, to be carried out below.
The main virtue of the toy model is its simplicity.
It will serve as a warm up exercise.

Let us assume that we have two complex fields $\phi$ and $\chi$, with 
one and the same electric charge, coupled to a U(1) gauge field (``photon"),
with the following self-interaction:
\beqn
{\cal L}_{\rm toy} = -\frac{1}{4 e^2}\, F_{\mu\nu}F^{\mu\nu} + \left| D_\mu\phi
\right|^2
+ \left| D_\mu\chi
\right|^2- V(\phi, \chi)\,,\nonumber\\[3mm]
V(\phi, \chi) =\frac{\lambda}{2}
\left\{\left(\bar\phi \phi - v^2\right)^2 + \left(\bar\chi \chi - v^2\right)^2
\right\}+ \beta |\phi |^2 |\chi |^2\,,
\label{syfone}
\eeqn
where $v$, $\lambda$ and $\beta$ are positive constants, and we assume, for simplicity,
that $\lambda \ll \beta\ll e^2\ll 1$.
It is easy to see that the model under consideration has two distinct minima
(classical vacua):

(i) $\phi$ develops a vacuum expectation value, $\chi$ does not;

(ii) $\chi$ develops a vacuum expectation value, $\phi$ does not.

\noindent
In the first case $|\phi | = v$, and one can always take
$\phi$ to be real and positive (this is nothing but imposing a gauge condition),
$\phi  = v$. The phase of $\phi$ is eaten up by the vector field,
which becomes massive, with   mass $m_V = \sqrt 2 \, ev$. The $\chi^\pm$ quanta
have mass $m_\chi = \sqrt {(\beta -\lambda )}\, v$. Finally, there is one real field
which remains from $\phi$; it can be parametrized as
$\phi = v +\eta$ with real $\eta$. The mass term of the $\eta$ field is
$2\lambda v^2\eta^2$, so that $m_\phi = \sqrt{ 2\lambda}\, v$.
In the second vacuum the mass of the vector field remains the same
while the roles of the $\phi$ and $\chi$ fields interchange, as well as their masses.
Note, however, that in neither vacuum there are massless excitations
which would make the vacuum manifold continuous. 
The energies in these two vacua are necessarily degenerate
because of the $Z_2$ symmetry $\phi\leftrightarrow\chi$
apparent in Eq. (\ref{syfone}) which is spontaneously broken.
Therefore, there must exist 
a {\em bona fide} domain wall interpolating between   vacua (i) and (ii).

Although the analytic solution for the domain wall seems to be unknown in
the case at hand, it is not difficult to
analyze its qualitative features.
Let us assume, for definiteness, that the wall lies in the 
$xy$ plane and impose the following boundary condition
(to be referred to as ``standard"):
\beq
\phi \to v, \,\,\, \chi\to 0\,\,\, \mbox{at}\,\,\, z\to-\infty ;
\qquad \chi \to v, \,\,\, \phi\to 0\,\,\, \mbox{at}\,\,\, z\to\infty \,.
\label{syftwo}
\eeq
Denote the ``standard" domain wall solution (centered at $z=0$),
with the above boundary conditions, as
\beq
\phi_0 (z) , \quad \chi_0 (z)\,.
\label{syfthree}
\eeq
Then 
\beq
 e^{i\sigma /2} \phi_0 (z-z_0) , \quad e^{-i\sigma /2} \chi_0 (z-z_0) 
\label{syffour}
\eeq
is obviously a solution too; it
represents a family of solutions with the same energy, containing two moduli
--- the wall center $z_0$ and a phase $\sigma$.
The occurrence of $z_0$ is due to the (spontaneous) breaking of the translational
invariance by the given wall solution, while $\sigma$ is due to
 the (spontaneous) breaking of a global U(1).

Indeed, the model (\ref{syfone}) has two   U(1) symmetries:
\beq
\phi \to \phi e^{i\gamma}\,,\quad \chi\to\chi\,;\quad\mbox{and}\quad
\phi \to \phi \,,\quad \chi\to\chi e^{i\beta}\,.
\label{syffive}
\eeq
One of these U(1) --- the diagonal combination --- is gauged, the other remains global.
It is not spontaneously broken in either of the vacua, (i) or (ii).
It is broken, however, on the domain wall.

\begin{figure}[h]  
\epsfxsize=7cm
\centerline{\epsfbox{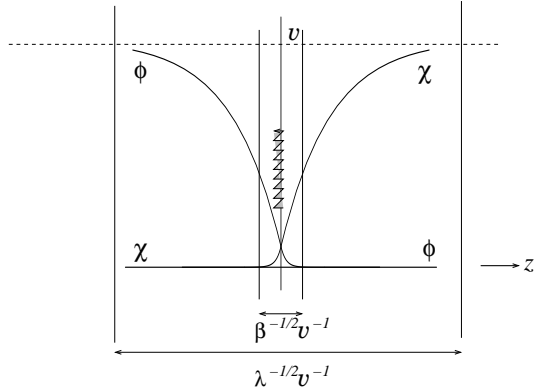}}
 \caption{A schematic rendition of the domain wall in the model (\ref{syfone}).}
\label{syfigone} 
\end{figure}

A qualitative sketch of the ``standard" domain wall
is given in Fig.~\ref{syfigone}.
Im$\phi$ = Im$\chi$=0 on the standard solution.
The field Re$\phi$ starts at $v$
at $z\to-\infty $, while Re$\chi$ starts at zero.
Then Re$\phi$ decreases as $v(1-e^{\sqrt\lambda v z})$ and 
Re$\chi$ increases as $v e^{\sqrt\beta v z}$. There is a crossover at $z=0$,
where the roles of Re$\phi$ and Re$\chi$ interchange: the field which 
was lighter becomes heavier, and {\em vice versa}. 
If one's goal is calculation of the wall tension,
one may treat the moduli $z_0$ and $\sigma$ as constants.
Then the vector field $A_\mu$ is not excited, $A_\mu =0$. 
The wall has a two-component 
structure: the thickness of one component of the wall is
$1/(\sqrt\lambda v)$ while that of the other $1/(\sqrt\beta v)$.
The latter size is much smaller than the former provided
that $\beta\gg\lambda$, as was assumed. The wall tension 
$T_{\rm w}$ is saturated by the contribution of the second
(narrow) component,
\beq
T_{\rm w} \sim v^3\sqrt\beta\,.
\label{syfsix}
\eeq

Our task is more than just calculating the wall tension.
We want to construct an effective 2+1 dimensional field theory
for the moduli on the wall world volume. As we will see momentarily,
to this end we will have to take into account the third
component of the wall, built of the gauge field,
which has thickness of order of $1/(ev)$ (see the zigzagy line in Fig. 
\ref{syfigone}). This component does not show up in the calculation of
$T_{\rm w}$.

Upon quantization, the moduli $z_0$ and $\sigma$ become fields
(adiabatically) depending on $x_m \equiv \{t,x,y\}$.
We will call them $\zeta (t,x,y )$ and $\sigma (t,x,y) $, respectively.
Furthermore, in Eq. (\ref{syffour}) the collective coordinates
$z_0$ and $\sigma$  are to be replaced by
\beq
z_0\longrightarrow \zeta (t,x,y )\,,\qquad \sigma \longrightarrow\sigma (t,x,y) \,.
\eeq
 Then we substitute
 the expressions for $\phi$ and $\chi$ in the Lagrangian
(\ref{syfone}), integrate over $z$ and obtain in this way
 a low-energy Lagrangian for the moduli.
For $\zeta (t,x,y )$ this procedure is quite standard, leading to
\beq
{\cal L}_\zeta = \frac{T_{\rm w}}{2}\left( \partial_m \zeta\right) \left( \partial^m \zeta\right) \,,
\qquad m =0,1,2.
\label{syfseven}
\eeq
For $\sigma (t,x,y) $ there is a subtlety. Indeed, substituting
Eq. (\ref{syffour}) in Eq. (\ref{syfone}), we arrive at
\beq
\bar\phi_0\phi_0\left|\frac{1}{2}\partial_m \sigma(t,x,y ) - A_m\right|^2
+\bar\chi_0\chi_0\left|\frac{1}{2}\partial_m \sigma(t,x,y ) + A_m\right|^2\,.
\label{syfeight}
\eeq
It is clear that at $z\to\pm\infty$ one must choose $A_m$ as follows:
\beq
A_m(t,x,y,z) \to \left\{\begin{array}{c}
\frac{1}{2}\, \partial_m \sigma(t,x,y )\quad{\rm at}\quad z\to -\infty\,,\\[3mm]
-\frac{1}{2}\, \partial_m \sigma(t,x,y )\quad{\rm at}\quad z\to \infty\,,
\end{array}
\right.
\label{syfnine}
\eeq
i.e.   $A_m$ is pure gauge. However, at $|z| \lsim (ev)^{-1}$
the field $A_m(t,x,y,z)$ must smoothly interpolate between
two regimes in Eq. (\ref{syfnine}) and, hence, cannot be pure
gauge in this domain of $z$ (see Fig. \ref{syfigone}).
It must be chosen in such a way as to minimize the coefficient
of $(\partial_m \sigma )^2$ in ${\cal L}_\sigma$. Thus, at $|z| \lsim (ev)^{-1}$
the photon field strength tensor is generated, with
necessity,
\beq
F_{m3} \sim \frac{\partial_m \sigma}{\Delta z}\,,\qquad \Delta z = (ev)^{-1}\,.
\label{syften}
\eeq
Therefore,
\beq
{\cal L}_\sigma = \kappa\, \frac{v}{e}\left( \partial_m \sigma \right) 
\left( \partial^m \sigma\right) ,
\label{syfeleven}
\eeq
where $\kappa$ is a numerical coefficient of order 1
depending on the form of $A_m$ in the intermediate domain.
It is determined through minimization.
Dynamics of the moduli fields on the wall world sheet
is thus described by the Lagrangian
\beq
{\cal L}_{2+1}= \kappa\, \frac{v}{e}\left( \partial_m \sigma \right) 
\left( \partial^m \sigma\right)  
+\frac{T_{\rm w}}{2}\left( \partial_m \zeta\right) \left( \partial^m \zeta\right) .
\label{syftwelve}
\eeq
The target space of the field $\sigma$ is $S_1$,
$$
0\leq\sigma\leq 2 \pi\,.
$$
As was noted by Polyakov \cite{Polyakov:1976fu},
in 2+1 dimensions, gauge field is equivalent to
a compact scalar field, through the relation
\beq
F_{mn}^{(2+1)} = \frac{e_{2+1}^2}{2\pi}\, \varepsilon_{mnk}\, \partial^k\sigma\,,
\label{syfthirteen}
\eeq
where $e_{2+1}^2$ is 2+1 dimensional gauge coupling (in our case
$e_{2+1}^2/(8\pi^2) =\kappa v/ e$). As a result, the moduli
Lagrangian (\ref{syftwelve})
can be rewritten as
\beq
{\cal L}_{2+1}= -\frac{1}{4 e_{2+1}^2}\, F_{mn}F^{mn} + 
\frac{T_{\rm w}}{2}\left( \partial_m \zeta\right) \left( \partial^m \zeta\right) .
\label{syffourteen}
\eeq

The consideration above bears purely illustrative character,
first and foremost because it was purely classical.
The impact of quantum corrections will be discussed in supersymmetric setting.

\section{Gauge invariant definition of the phase collective coordinate
$\sigma$}
\setcounter{equation}{0}

When we speak of the relative phase of the fields 
$\phi$ and $\chi$ we compare their phases
in distant points, $z\to\pm\infty$.
In the theory with the local gauge invariance this
raises the question as to the
meaningfulness of this comparison. The way we have introduced the
phase collective coordinate
$\sigma$ in Sect. \ref{toymodel} is meaningful only in a specific gauge.
In fact, it is useful to give a gauge invariant definition.
Let us introduce $\sigma$ as follows:
\beq
\sigma = {\rm Arg}\,\phi (z_1) - {\rm Arg}\,\chi (z_2) 
+\int_{z_1}^{z_2}\, dz\, A_3 (z)\,,
\label{dids}
\eeq
where formally $z_1\to -\infty\,, \,\, z_2 \to \infty $.
Under the gauge transformations
\beq
\phi (z) \to e^{i\alpha (z)}\phi (z) ,\quad \chi (z) \to e^{i\alpha (z)}\chi (z) ,
\quad A_3 (z) \to A_3 (z) +\partial_z \alpha (z)\,,
\label{gtfpac}
\eeq
while $\sigma$ as defined in Eq. (\ref{dids}) stays intact.

Although formally $z_1\to -\infty\,, \,\, z_2 \to \infty $,
there is an unquestionable tradition to set
$\phi ,\,\,\,\chi$ = const, $A_\mu=0$ in the plane vacuum
(this is the unitary gauge).
Therefore, practically one can take
$z_1$ just to the left of the wall while $z_2$ just to the right.
With this choice of $z_{1,2}$ it is perfectly clear
that $\sigma$ is a collective coordinate characterizing the
wall structure, an internal phase. 

Alternatively one can consider a (nonlocal
gauge invariant)  order parameter
\beq
\langle\, \bar\chi (z_2) \, e^{i\int_{z_1}^{z_2}\, dz\, A_3 (z)}\,\phi (z_1)\,\rangle .
\label{nlopgi}
\eeq 
It obviously vanishes in the plane vacuum,
while it reduces to $v^2 e^{i\sigma }$ if there is a domain wall
of the type discussed in Sect. \ref{toymodel}.
This order parameter is noninvariant under the ``second" U(1) ---
the one which is not gauged. The fact that (\ref{nlopgi})
does not vanish 
on the wall means that the global U(1)
is spontaneously broken triggering the emergence of the
Goldstone boson localized on the wall and described
by the field $\sigma (t,x,y )$.

\section{Basic model}
\label{basicmodel}
\setcounter{equation}{0}

\subsection{The original non-Abelian theory}
\label{original}

The field content
of \ntwo SQCD with the gauge group SU(2) and $N_f$ flavors of the quark multiplets
is well-known. The \ntwo vector multiplet consists of the gauge field $A_{\mu}^a$, two
Weyl fermions $\lambda^{fa}_{\alpha}$ and the scalar field $\phi^a$, 
all in the adjoint representation of the gauge group. Here  $a=1,2,3$ is the color index,
 $\alpha=1,2$
is the spinor index  and $f=1,2$ is an SU(2)$_R$ index
(warning: SU(2)$_R$ is not to be confused   with the gauge SU(2) or flavor SU(2)
emerging at $N_f=2$, the case to be considered below).

 On the Coulomb branch, the adjoint scalar develops a VEV,
\beq
\langle \phi^3\rangle \equiv  \langle  a \rangle 
\label{gsbc}
\eeq 
breaking the gauge group down to  U(1). The W bosons and their
superpartners become heavy (with masses of order of $\langle  a \rangle
\sim m_A\gg \Lambda$) and 
can be integrated out. What is left of the vector multiplet in the low-energy SQED
are the  third color components of $A$, $\lambda$ and $\phi$,
to be referred as $A_{\mu}$, two
Weyl fermions $\lambda^{f}_{\alpha}$,  and the scalar field $a$.
This is the field content of SQED.

The quark multiplets of the non-Abelian theory consist of   complex scalar fields
$q^{kA}$ and $\tilde{q}_{Ak}$ and   Weyl fermions $\psi^{kA}$ and $\tilde{\psi}_{Ak}$,
 all in the fundamental representation of the gauge group.
Here
 $k=1,2$ is the color index
while $A$ is the flavor index. In what follows we will limit ourselves
to $A=1,2$. 

With the
  gauge symmetry spontaneously broken by
the condensate (\ref{gsbc}), only the upper ($k=1$) 
components, or only the lower ($k=2$) components
 of the  fields $q, \,\, \tilde{q}, \,\, \psi$ and $\tilde{\psi}$
remain light in the quark vacua,
 while the opposite components acquire the same mass as the W bosons
and the other fields in the \ntwo vector supermultiplet.
They
can be integrated out
and will play no role in our consideration. 

Let us denote
the light quark  fields as 
$q^{A}$, $\tilde{q}_{A}$ and $\psi^{A}$, and $\tilde{\psi}_{A}$, respectively.
Note that the scalars form a doublet under the action of global  SU(2)$_R$ group,
$q^f=(q,\bar{\tilde{q}})$.
In terms of these fields the bosonic part of the low-energy effective SQED
takes the form \footnote{Here and below we use a
formally  Euclidean notation, e.g.
$F_{\mu\nu}^2 = 2F_{0i}^2 + F_{ij}^2$,
$\, (\partial_\mu a)^2 = (\partial_0 a)^2 +(\partial_i a)^2$, etc.
This is appropriate as long as we after static (time-independent)
field configurations, and $A_0 =0$. Then the Euclidean action is
nothing but the energy functional. Furthermore, we
 define $\sigma^{\alpha\dot{\alpha}}=(1,-i\vec{\tau})$,
 $\bar{\sigma}_{\dot{\alpha}\alpha}=(1,i\vec{\tau})$. Lowing and raising of spinor indices
is performed by 
virtue of the antisymmetric tensor defined as $\ve_{12}=\ve_{\dot{1}\dot{2}}=1$,
 $\ve^{12}=\ve^{\dot{1}\dot{2}}=-1$. }
\begin{eqnarray}
S_{\rm low-en} &=&\int d^4 x \left\{ \frac1{4 \gs}F_{\mu\nu}^2 +
\frac1{\gs} \left|\partial_{\mu}a\right|^2
+\bar{\nabla}_{\mu}\bar{q}_A\nabla_{\mu}q^A+
\bar{\nabla}_{\mu}\tilde{q}_A\nabla_{\mu}\bar{\tilde{q}}^A \right.
\nonumber\\[3mm]
&+&\frac{g^2}{8}\left( |q^A|^2 -|\tilde{q}_A|^2\right)^2 +\frac{g^2}{2}\left|\qt_{A}q^A-\frac{f ( a)}{2}\right|^2
\nonumber\\[3mm]
&+&\left.\frac12 \left( |q^A|^2+|\tilde{q}_A|^2\right)\, \left|a+\sqrt{2}m_A\right|^2
\right\},
\label{qed}
\end{eqnarray}
where $\nabla_{\mu}=\partial_{\mu} -\frac{i}{2}A_{\mu}$, 
$\bar{\nabla}_{\mu}=\partial_{\mu} +\frac{i}{2}A_{\mu}$, while 
near the quark vacua\,\footnote{In our case the variable $u$ defined
in Eq.~(\ref{brsup}) can be represented as
$u=a^2/2=(1/2)\, (-\sqrt 2 m +\delta a)^2.$ Hence $f(a)= -2\sqrt{2}\, \mu\,  
 \partial u/\partial a =4\mu m -2\sqrt 2\mu\delta a.$}
\beq
f(   a ) = \xi -2\sqrt 2\,  \mu\, \delta a + ...
\eeq
and the ellipses denote terms quadratic in $\delta a$.
Under our choice of parameters  the generalized
  FI parameter $\xi$ can be chosen as 
\beq
\label{xim}
\xi=4\,\mu\,m,
\eeq
where 
\beq
m=\frac12(m_1+m_2)\,.
\label{defavm}
\eeq
 In order to keep $\xi$ real (as we will always do)
we assume that
both $\mu$ and $m$ are real.
The scalar potential in Eq. (\ref{qed}) comes from $F$ and $D$ terms of the vector and
matter supermultiplets.

\subsection{\ntwo SQED}
\label{n2sqed}

In what follows we will mostly ignore the $a$ dependence of
the function $f(a)$ in the second line of Eq.~(\ref{qed})
setting $f(   a ) = \xi $. The theory we get in this limit 
is \ntwo SQED. The relative impact of
the terms linear in $\delta a$ in the function $f(a)$ is of order
$m^{-1}\,\delta a$. As will be seen momentarily, $\delta a \sim m_2-m_1$.
Therefore,  taking   account of the
dependence of $f(a)$ on $a$
results in a small (\ntwo breaking) correction
provided $|\Delta m|\ll m$, as we always assume. All our conclusions 
remain unchanged.

From Eq.~(\ref{qed}) we can immediately infer the vacuum structure of the model at hand. 
We have two quark vacua, the first one located at
\beq
\label{a1}
a=-\sqrt{2}\,m_1
\eeq
 while the second one   at
\beq
\label{a2}
a=-\sqrt{2}\,m_2.
\eeq
In the first vacuum the first quark flavor develops a VEV,
\beq
\label{q1}
q^1=\qt_1=\sqrt{\frac{\xi}{2}},\qquad  q^2=\qt_2=0\,,
\eeq
completely breaking the U(1) gauge symmetry.  In the 
second vacuum the second flavor develops  a VEV,
\beq
\label{q2}
q^2=\qt_2=\sqrt{\frac{\xi}{2}},\qquad  q^1=\qt_1=0\,,
\eeq
which does the same job.

Above, the vacuum expectation values of the
squark fields were assumed to be real.
In fact, one can assign arbitrary phases to the two squark fields.
We will discuss the impact of these phases shortly.

Now let us discuss the mass spectrum of light fields in both quark vacua.
Consider for definiteness  the first vacuum, Eq.~(\ref{a1}).
 The spectrum  can be obtained by diagonalizing  the quadratic form
in  (\ref{qed}). This is done in Ref.~\cite{VY};  the result is as follows:
 one real component of field $q^1$ is eaten up by the Higgs mechanism
to become the third components of the massive photon. Three components
of the massive photon, one remaining component of $q^1$ and four real components
of  the fields $\qt_1$ and $a$ form one long \ntwo multiplet (8 boson states
+ 8 fermion states), with mass
\beq
\label{mgamma}
m_{\gamma}^2=\frac12\, g^2\,\xi.
\eeq

The second flavor $q^2$, $\qt_2$ (which does not condense in this vacuum)
forms one short \ntwo multiplet (4 boson states + 4 fermion states), with
mass $\Delta m$ which is heavier than
the mass of the vector supermultiplet\,\footnote{In estimating the
relative  masses of
various fields we will ignore the factors of $g^2$.
Although the coupling constant is certainly assumed to be small to
make the theory weakly coupled, we consider the $\mu, m$ dependences
as more important.}. The latter assertion
applies to the regime (\ref{mxi}).
In this regime the $W$-boson supermultiplet is heavier still.

If we consider the limit opposite to that in Eq.~(\ref{mxi}) and tend
$\Delta m\to 0$,
the ``photonic" supermultiplet becomes heavier than that of $q^2$,
the second flavor field.
therefore, it can be integrated out, leaving us with 
the theory of massless moduli from  $q^2$,
which interact through a nonlinear sigma model with the K\"ahler term
corresponding to 
the Eguchi-Hanson metric.

In the second vacuum the mass spectrum is   similar --- the 
roles of the first and the second flavors are interchanged.

Our immediate goal is
constructing 1/2 BPS domain wall interpolating between the above two vacua.

\section{The domain wall in \ntwo SQED}
\label{domainwall}
\setcounter{equation}{0}

In this section we work out and solve the first order 
Bogomolny equations for 
 the domain wall. The Bogomolny equations
can be derived in two ways:
by performing the Bogomolny
completion \cite {B} and by analyzing
the set of supercharges and isolating those four that
annihilate the wall \cite{OW,deAzcarraga:gm,DS,HS,DDT,GS}. We will follow both routes.

\subsection{First-order equations}
\label{firstorder}

 First,  let us note that the structure of the vacuum
condensates in both vacua (\ref{q1}) and (\ref{q2}) suggests that we can
look for domain wall solution using the {\em ansatz}
\beq
\label{qqt}
q^A=\bar{\qt}_A\equiv \frac{1}{\sqrt{2}}\,\vp^A\, ,
\eeq
where we introduce a new complex field  $\vp^A$.
In this ansatz SQED under consideration reduces to 
\begin{eqnarray}
S_{\rm red \,\, SQED} &=&\int d^4 x \left\{ \frac1{4\gs}F_{\mu\nu}^2 
+\frac1{\gs}|\partial_{\mu}a|^2
+\bar{\nabla}_{\mu}\bar{\vp}_A\nabla_{\mu}\vp^A
\right.
\nonumber\\[3mm]
&+&\left. \frac{g^2}{8}\left( |\vp^A|^2 -\xi \right)^2
+\frac12 \left|\vp^A\right|^2\, \left| a+\sqrt{2}m_A\right|^2
\right\}\,.
\label{redqed}
\end{eqnarray} 
This Lagrangian is qualitatively very similar to that considered
in Sec. \ref{toymodel}.
In particular, it has two {\em global} U(1) symmetries, allowing one
to independently rotate
the fields of the first and second flavors, respectively.
The diagonal U(1) is gauged. 

In terms of the fields $\vp^A$ the quark condensate becomes
\beq
\label{p1}
\vp^1=\sqrt{\xi}\, \exp (i \alpha ) \,,\qquad \vp^2=0
\eeq
in the vacuum (\ref{a1}), and
\beq
\label{p2}
\vp^2=\sqrt{\xi}\, \exp (i\alpha ' )\,,\qquad \vp^1=0
\eeq
in the vacuum (\ref{a2}). Because of the gauge freedom, the phases $\alpha$, $\alpha '$
can be always chosen to vanish.    

If we  assume that all fields depend only on the
coordinate $z=x_3$
the Bogomolny completion of the wall energy functional can be written as
\begin{eqnarray}
T_{\rm w} &=& \int dz \left\{ \left|\nabla_z \vp^A\pm \frac1{\sqrt{2}}\vp^A(a+
\sqrt{2}m_A)\right|^2
\right.
\nonumber\\[3mm]
\label{bog}
&+& \left.\left|\frac1{g}\pz a \pm \frac{g}{2\sqrt{2}}(|\vp^A|^2-\xi)\right|^2
\pm\frac1{\sqrt{2}}\xi\pz a\right\}.
\end{eqnarray}
Here we do not assume 
{\em a priori}
that the gauge field vanishes. We will see that although the
gauge field strength does vanish on the flat wall solution at rest, the gauge
potential need not vanish.
Putting  mod-squared terms   to zero gives us the first order Bogomolny
equations,  
while the surface term (the last one in Eq.~(\ref{bog}))
gives the wall tension. Assuming for definiteness that
$\Delta m>0$ and choosing the upper sign in (\ref{bog}) we get the BPS 
equations,
\begin{eqnarray}
\nabla_z \vp^A &=& -\frac1{\sqrt{2}}\vp^A\left(a+\sqrt{2}m_A\right),
\nonumber\\[3mm]
\label{wfoe}
\pz a &=&- \frac{g^2}{2\sqrt{2}}\left(|\vp^A|^2-\xi\right).
\end{eqnarray}
These first order equations should be supplemented by the following boundary 
conditions
\begin{eqnarray}
\vp^1(-\infty) &=&\sqrt{\xi},\quad\vp^2(-\infty)=0, \quad a(-\infty)=-\sqrt{2}m_1\,;
\nonumber\\[3mm]
\label{wbc}
\vp^1(\infty) &=& 0,\quad \vp^2(\infty)=\sqrt{\xi},\quad a(\infty)=-\sqrt{2}m_2,
\end{eqnarray}
which show  that our wall interpolates between the two quark vacua.
As was mentioned, this boundary condition is not generic
(it was referred to as ``standard" in Sec. \ref{toymodel}).
The existence of the exact ungauged U(1)  implies that the generic boundary condition
could be obtained from Eq. (\ref{wbc}) by multiplying $\phi^1$ in the first line
by $e^{i\sigma/2}$ and  $\phi^2$ in the second line
by $e^{-i\sigma/2}$.

Finally, the tension of the wall satisfying the above equations is
\beq
\label{wten}
T_{\rm w}\,=\,|( \Delta m )\,  \xi|.
\eeq

Now, let us derive the Bogomolny equations by analyzing relevant
combinations of supercharges. We will show that four combinations
of supercharges act trivially.
To see this explicitly,
let us write down the supersymmetry transformations in  
SQED:
\begin{eqnarray}
&&
\delta\lambda^{f\alpha}=\frac12(\sigma_\mu\bar{\sigma}_\nu\ve^f)^\alpha
F_{\mu\nu}+\ve^{\alpha p}D^a(\tau^a)^f_p
+i\sqrt{2}\partial\hspace{-0.65em}/^{\,\,\alpha\dot{\alpha}}\, a \, 
\bar{\ve}^{f}_{\dot\alpha}\ ,
 \nonumber\\[2mm]
&& \delta\psi^{\alpha A}\ =\ i\sqrt2\
\nabla\hspace{-0.65em}/^{\,\,\alpha\dot{\alpha}}q^{fA}\bar{\ve}_{f\dot{\alpha}}
+\sqrt{2}\ve^{\alpha f}F_f^{A}\ ,
\nonumber\\[2mm]
&& \delta\tilde{\psi}^{\alpha}_{A}\ =\ i\sqrt2\
\nabla\hspace{-0.65em}/^{\,\,\alpha\dot{\alpha}}\bar{q}^f_{A}\bar{\ve}_{f\dot{\alpha}}
+\sqrt{2}\ve^{\alpha f}\bar{F}_{fA}\ ,
\label{str}
\end{eqnarray}
where we explicitly write out the SU(2)$_R$ indices $f,p=1,2$.
Here $D^a$ is the SU(2)$_R$ triplet of $D$ terms which in the {\em ansatz}
(\ref{qqt}) reduce to
\beq
\label{dterm}
D^1=i\frac{g^2}{2}\left(|\vp^A|^2-\xi\right),\; D^2=D^3=0\,,
\eeq
while $F^f$ and $\bar{F}_f$ are the matter $F$ terms,
\beq
\label{fterm}
F^{fA}=i\frac1{\sqrt{2}}\left(a+\sqrt{2}m_A\right)q^{fA},\qquad
\bar{F}_{Af}=i\frac1{\sqrt{2}}\left(\bar{a}+\sqrt{2}m_A\right)\bar{q}_{Af}.
\eeq

The fact that the wall we work with is critical implies that
some of
$\delta\lambda$ and  $\delta\psi$ in Eq.~(\ref{str}) vanish.
Accepting, as above, the {\em ansatz}
 (\ref{qqt}) and taking into account that all fields depend only on $z$
we get the same first order equations (\ref{wfoe}), provided
that $\ve^{\alpha f}$ and $\bar{\ve}_{\dot{\alpha}}^f$
satisfy the following conditions
\begin{eqnarray}
\bar{\ve}^2_{\dot{2}}&=&-i\ve^{21},\quad
\bar{\ve}^1_{\dot{2}}=-i\ve^{22},
\nonumber\\[3mm]
\label{wepsilon}
\bar{\ve}^1_{\dot{1}}&=&i\ve^{12},\qquad
\bar{\ve}^2_{\dot{1}}=i\ve^{11}.
\end{eqnarray}
These four constraints on the supertransformation
parameters show which particular linear combinations of
the supercharges act  trivially on the domain wall solution.
With these four constraints we  reduce  the number of 
trivially acting supercharges   to four 
(out of eight). Thus, our domain wall is 1/2 BPS-saturated.

\subsection{Finding the domain wall solution}
\label{wallsolution}

Now let us work out the solution to the first order equations (\ref{wfoe}),
assuming the conditions (\ref{mxi}) to be satisfied. The range of variation of 
the field
$a$ inside the wall is of the order of $\Delta m$ (see Eq.~(\ref{wbc})). 
The  minimization of  its kinetic energy
implies this field to be slowly varying.
Therefore, we may safely assume that the wall is thick; its  size $R\gg 1/\sqrt{\xi}$.
This fact will be confirmed shortly.

\begin{figure}[h]  
\epsfxsize=9cm
\centerline{\epsfbox{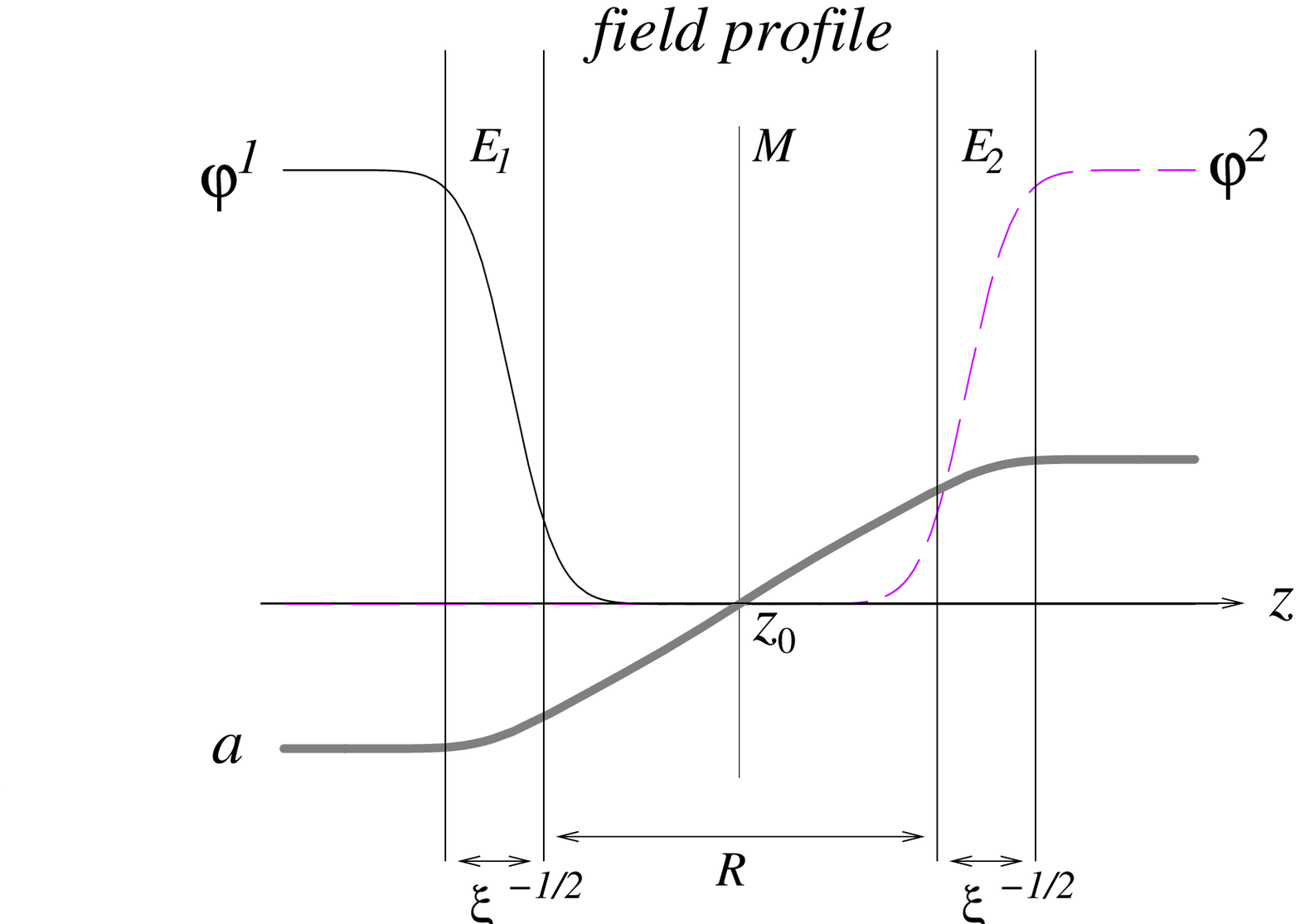}}
 \caption{Internal structure of the domain wall:
two edges (domains $E_{1,2}$) of the width $\sim \xi^{-1/2}$
are separated by a broad middle band (domain $M$) of the width $R$,
see Eq. (\ref{R}).}
\label{syfigthree} 
\end{figure}

We arrive at the following picture of the domain wall at hand. The quark fields
vary from their VEV's $\sim\sqrt{\xi}$ to zero within small regions, of the order of
$1/\sqrt{\xi}$ (see the previous footnote). They remain small inside the wall, see Fig. 
 \ref{syfigthree}.

Then to the leading order we can put the quark fields to zero 
in (\ref{wfoe}).
Now, the second equation in (\ref{wfoe}) tells us that $a$ is a linear function
of $z$. The solution for $a$ takes the form
\beq
\label{a}
a=-\sqrt{2}\left( m-\Delta m \frac{z-z_0}{R}\right),
\eeq
where the collective coordinate $z_0$ is the position  of the wall center
(and $\Delta m$ is assumed positive). 
The solution  is valid in a
wide domain of $z$ 
\beq
\label{inside}
\left| z-z_0\right| < \frac{R}{2}\,,
\eeq
except narrow areas of size $\sim 1/\sqrt{\xi}$ near the edges of the wall
at $z-z_0=\pm R/2$.

Substituting the solution (\ref{a}) in the second equation in (\ref{wfoe})
we get
\beq
\label{R}
R=\frac{4\Delta m}{g^2 \xi}= \frac{2\Delta m}{m_\gamma^2}\,.
\eeq
Since $\Delta m /\sqrt\xi \gg 1$, see Eq. (\ref{mxi}), this result shows that $R\gg1/\sqrt{\xi}$,
 which justifies our approximation. It is easy to check that
1/2 of the wall tension quoted in Eq.~(\ref{wten}), comes from the kinetic term of the
field $a$ in the middle domain $M$.

Furthermore, we can now use the first relation in Eq.~(\ref{wfoe}) 
to determine   tails
of the quark fields inside the wall. First let us fix the gauge imposing 
the condition that $\vp^1$ is real at $z\to -\infty$ and 
$\vp^2$ is real at $z\to \infty$. This is a generalization of the unitary
gauge for the problem with domain wall that interpolates between two
vacua. Of course, this requirement does not fix the gauge completely.
We still have freedom to make gauge transformations  with gauge parameter
which is non-zero inside the wall.

Let us assume   that the gauge field is given by
\beq
\label{wgpot}
A_z=\sigma \, \pz \beta (z),\qquad A_n=0,\quad n=0,1,2,
\eeq
so that the field strength is zero. Here $\beta (z)$ is some function
of $z$ while $\sigma$ is a constant introduced in order to normalize
$\beta (z)$ in a convenient way.

Consider first the left edge (domain $E_1$ in Fig. \ref{syfigthree})
at
$z-z_0=-R/2$. Substituting the above solution for $a$ in the equation for
$\vp^1$ we get
\beq
\label{wq1}
\vp^1=\sqrt{\xi}\, e^{-\frac{m_{\gamma}^2}{4}\left(z-z_0+\frac{R}{2}\right)^2 
+i\frac{\sigma}{2}[1+\beta (z)]}\,\, .
\eeq
This behavior is valid in the domain $M$, at  $(z-z_0+R/2)\gg1/\sqrt{\xi}$, and shows 
that the field of the first quark flavor tends to zero  exponentially
inside the wall, as was expected. 
Our gauge choice requires
\beq
\label{betam}
\beta (z)\to -1,\quad  z\to -\infty,
\eeq
while inside the wall the function $\beta (z)$ remains undetermined
reflecting the possibility of gauge transformations.

By the same token,  we can consider the behavior of the second quark flavor
near the right edge of the wall at  $z-z_0=R/2$. The 
first of equation  in (\ref{wfoe})   for $A=2$
implies
\beq
\label{wq2}
\vp^2=\sqrt{\xi}\,
e^{-\frac{m_{\gamma}^2}{4}\left(z-z_0-\frac{R}{2}\right)^2 -i\frac{\sigma}{2}
[1-\beta (z)]}\,\,,
\eeq
which is valid in the domain $M$ provided that $(R/2-z+z_0)\gg1/\sqrt{\xi}$.
Inside the wall the second 
quark flavor tends to zero  exponentially too. Our gauge choice implies
that
\beq
\label{betap}
\beta (z)\to 1,\quad  z\to \infty.
\eeq

Needless to say that
the first and second quark flavor profiles are symmetric
with respect to reflection at $z_0$. The potential term of the $\varphi^A$ fields
in the domain $M$ produces the remaining 1/2 of the wall tension,
\beq
\int dz\,\, \frac{g^2}{8}\left(\left|\varphi^A\right|^2-\xi\right)^2 =\frac{g^2}{8}\,\xi^2\,R
=\frac{T_{\rm w}}{2}\,.
\label{rem12}
\eeq
This means of course that the contribution of the edges $E_{1,2}$ in the wall
tension must be of higher order in $\xi$. With this 
remark we proceed
to   the edge domains.

\vspace{1mm}

In the domains near the wall edges, $z-z_0=\pm R/2$, the fields $\vp^A$ and $a$
smoothly interpolate between their VEV's in the given vacua and the 
behavior inside the wall determined by Eqs.~(\ref{a}) and (\ref{wq1}), and (\ref{wq2}).
It is not difficult to check that these domains produce    contributions
to the wall tension of the order of $\xi^{3/2}$ which makes them negligibly small.

Now let us comment on the   phase factors in (\ref{wq1}), (\ref{wq2}).
Two complex fields $\vp^1$ and $\vp^2$ have, generally speaking,
 two independent phases. Since it is the diagonal U(1) that is gauged
it is natural to parametrize them as
 a common phase 
which we denote as $\beta (z)\sigma/2$ (it  depends on  the
gauge transformations which we can  still make inside the wall)
and a relative phase $\sigma$.   This phase  $\sigma$ is nothing but a 
global  U(1) remnant
 of the global flavor  SU$(N_f=2)$ symmetry
which is explicitly broken down to U(1)
due to the fact that $m_1\neq m_2$.
It is worth stressing that $\sigma$ is a {\em collective coordinate
of the wall}, rather than a modulus associated with the vacua. 
Because of the   U(1)$\times$U(1)
symmetry of the Lagrangian (\ref{redqed}), the effective theory on the wall
has no potential energy associated with $\sigma$.
At the same time, as was already mentioned,
in the model under consideration there are no massless fields in the bulk ---
 all fields are massive in each of two quark vacua.

Thus we have two collective coordinates characterizing our wall solution,
the position of the center $z_0$ and the phase $\sigma$. In the effective 
low-energy   theory on the wall
they become scalar fields of the world-volume 2+1 dimensional theory,
$\zeta (t,x,y)$ and $\sigma (t,x,y)$, respectively.
 The target space of the second field is
$S_1$, as is obvious from Eqs. (\ref{wq1}) and (\ref{wq2}).

\subsection{Zero modes }

\begin{figure}[h]  
\epsfxsize=7cm
\centerline{\epsfbox{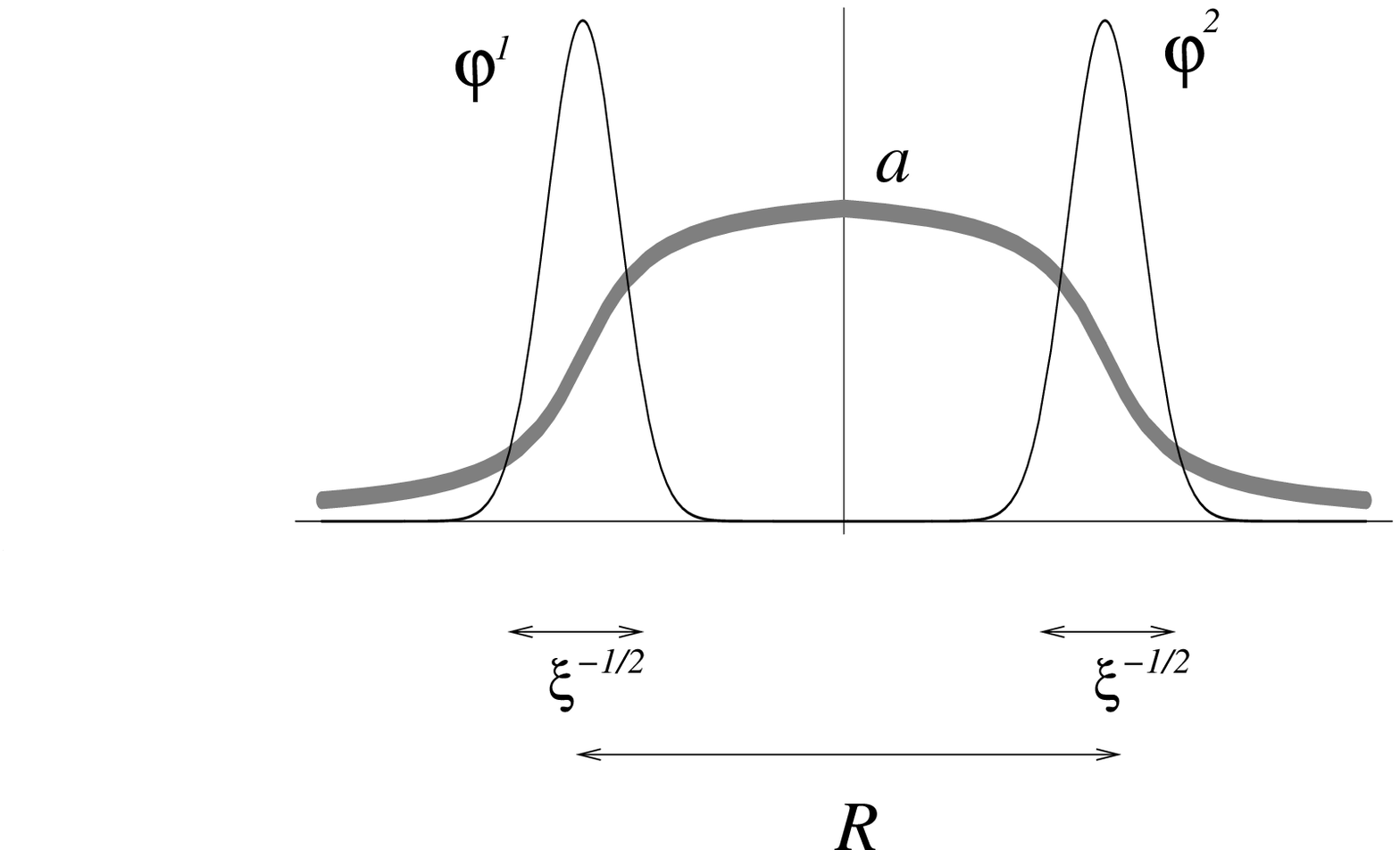}}
 \caption{The translational zero mode.}
\label{syfigfour} 
\end{figure}

Two bosonic zero modes are obtained by differentiating
 the solution of  Sec. \ref{wallsolution}
with respect to $z_0$ and $\sigma$. The first one is translational, the second 
can be called
``rotational." Obtaining the translational zero mode is straightforward; it is depicted
in Fig. \ref{syfigfour}, which gives an idea of its spread
(localization). The rotational zero mode deserves a comment.

The solution for the $a$ field is $\sigma$ independent. Therefore
the rotational zero mode  contains no $a$ component.
As for the $\phi$ components, differentiating the solution with
respect to  $\sigma$ in a straightforward manner
one obtains 
\beq
\vp^A = \phi_0^A +\vp_{\rm zm}^A\,,\qquad A=1,2\,,
\eeq
where
\begin{eqnarray}
\vp_{\rm zm}^1&=&\frac{\partial\vp_0^1}{\partial\sigma}\delta\sigma 
=\frac{i}{2} \phi_0^1 [1+\beta(z)]\delta\sigma \,,
\nonumber\\[3mm]
\vp_{\rm zm}^2&=&\frac{\partial\vp_0^2}{\partial\sigma}\delta\sigma 
=-\frac{i}{2}  \vp_0^2 [1-\beta(z)]\delta\sigma .
\end{eqnarray}
Note that because of the boundary conditions (\ref{betam}), (\ref{betap})
these zero modes are normalizable.
Qualitatively
they are the same as the $\vp$ modes in Fig. \ref{syfigfour}.

Now we will dwell on the fermion zero modes.
There are fermion zero modes of two types, ``supertranslational" and
``super-rotational."

To generate these modes we apply supersymmetry transformations to the bosonic
wall solution of Sec. \ref{wallsolution}. As we 
have already explained, four supercharges selected by  
conditions (\ref{wepsilon}) act trivially on the wall solution. The remaining four
supercharges act non-trivially giving fermion zero modes.
To separate them we impose conditions between $\ve^{\alpha f}$ and
$\bar{\ve}_{\dot{\alpha}}^{f}$ which differs by sign from those in 
Eq.~(\ref{wepsilon}).
Namely, we take
\begin{eqnarray}
\bar{\ve}^2_{\dot{2}}&=&i\ve^{21},\quad
\bar{\ve}^1_{\dot{2}}=i\ve^{22},
\nonumber\\[3mm]
\label{wzepsilon}
\bar{\ve}^1_{\dot{1}}&=&-i\ve^{12},\qquad
\bar{\ve}^2_{\dot{1}}=-i\ve^{11}\,,
\end{eqnarray}
and substitute this into Eq.~(\ref{str}). Using (\ref{wfoe}) we get
the fermion zero modes on the wall
\begin{eqnarray}
\lambda^{\alpha f}&=& i\eta^{\alpha p}(\tau_1)^f_p 
g^2\left(|\vp^A|^2-\xi\right),\qquad f,p =1,2,
\nonumber\\[3mm]
\psi^{\alpha A} &=&  i\eta^{\alpha f}\left(a+\sqrt{2}m_A\right)q_f^{A},
\nonumber\\[3mm]
\tilde{\psi}^{\alpha}_A &=& i\eta^{\alpha f}\left(a+\sqrt{2}m_A\right)
\bar{q}_{fA},
\label{wfzeromodes}
\end{eqnarray}
where by definition
\beq
\label{qbarq}
q_f^{A}=\frac1{\sqrt{2}}(\vp^A,\, -\vp^A),\quad
\bar{q}_{fA}=\frac{1}{\sqrt{2}}(\bar{\vp}_A,\, \bar{\vp}_A)\,,
\eeq
while $\eta^{\alpha f}$ are four Grassmann parameters parametrizing
the fermion zero modes.
Under the supertransformations
 $$\delta\eta^{\alpha f}=\ve^{\alpha f}.$$
Here the fields $a$ and $\vp^A$ are given by Eqs.~(\ref{a}) and (\ref{wq1}),
 (\ref{wq1}).
Note that the conjugated fermion fields are given by conjugation 
of Eqs.~(\ref{wfzeromodes}), however the parameters
 $\bar{\eta}_{\dot{\alpha}}^ f$ are not independent. They are expressed 
in terms of $\eta^{\alpha f}$ according to
\begin{eqnarray}
\bar{\eta}^2_{\dot{2}}&=&i\eta^{21},\quad
\bar{\eta}^1_{\dot{2}}=i\eta^{22},
\nonumber\\[3mm]
\label{bareta}
\bar{\eta}^1_{\dot{1}}&=&-i\eta^{12},\qquad
\bar{\eta}^2_{\dot{1}}=-i\eta^{11}.
\end{eqnarray}

We see that the $\lambda$ fermions (i.e. photino plus its \ntwo partner) are non-zero 
and approximately
constant inside the wall while the matter fermions $\psi$ are
concentrated near the wall edges. This picture is a super-reflection of that
in Fig. \ref{syfigfour}.

\subsection{Effective field theory on the wall}
\label{kinterms}

In this subsection we work out (2+1)-dimensional  effective low-energy
theory 
of the moduli on the wall. To do so we make the wall collective coordinates
$z_0$ and  $\sigma$ (together with their fermionic superpartners $\eta^{\alpha f}$)
slowly varying fields depending on 
 $x_n$  ($n=0,1,2$),
$$
z_0 \to \zeta (x_n)\,,\quad \sigma \to \sigma  (x_n)\,,\quad
\eta^{\alpha f}\to \eta^{\alpha f}(x_n)\,.
$$
 For simplicity let us consider the
bosonic fields $\zeta (x_n)$ and  $\sigma  (x_n)$;
the  residual supersymmetry will allow us to
readily reconstruct the fermion part of the effective action.

Because $\zeta (x_n)$ and  $\sigma  (x_n)$ correspond to zero modes of the wall,
 they have no  
potential terms in the world sheet theory. Therefore, in fact our task is to derive
kinetic terms. 
For $\zeta (x_n)$ this procedure is very simple. Substituting
the wall solution (\ref{a}), (\ref{wq1}), and (\ref{wq2}) in  the action
(\ref{redqed}) and taking into account the $x_n$ dependence of this modulus
 we immediately get (cf. (\ref{syfseven}))
\beq
\label{kinz0}
\frac{T_{\rm w}}{2} \, \int d^3 x \; (\pt_n \zeta   )^2\, .
\eeq
As far as the kinetic term for $\sigma  (x_n)$ 
is concerned more effort is needed. We start from 
Eqs. (\ref{wq1}) and  (\ref{wq2}) for the quark fields\,\footnote{
Strictly speaking, these equations {\em per se}
are valid only inside the wall, in the domain $M$.
Outside the wall $\vp^1\to \sqrt{\xi}\exp [(i\sigma/2)(1+\beta (z))]$
at $z\to -\infty$
and $\vp^2\to \sqrt{\xi}\exp [(-i\sigma/2)(1-\beta (z))]$
at $z\to \infty$.}. 
Then we will have to modify our {\em ansatz}
for the gauge field\,\footnote{Remember the 
electric charge of the quark fields is $\pm 1/2$.},
\beq
\label{kingpot}
A_\mu=\pt_{\mu}[\sigma(x_n)  \beta (z)]+\chi(z)\, \pt_{\mu}\sigma(x_n)\,.
\label{nepf}
\eeq
A few important points to be noted are as follows.

(i) We have  introduced an extra  profile function  $\chi(z)$.
It has no role in the construction of the static wall solution
{\em per se}. It is unavoidable, however, 
in constructing the kinetic part of the world sheet theory
of the moduli.
This new profile function
 is described by its own action, which will be subject to
minimization. This seems to be an element of the procedure
which is sufficiently general (previously a similar construction
was applied e.g. in Ref. \cite{Shifman:2002yi}), and yet, to the best of our knowledge,
  no  proper  coverage can be found in the literature. 

(ii)  The first term in Eq. (\ref{nepf}) is   pure gauge;
it replaces a similar term 
in Eq.  (\ref{wgpot}) where  $\sigma$ was  $x$ independent. 
The function $\beta (z)$ remains undetermined, and so does
the pure gauge term. This should not worry us since
they do not show up in any physical observables.
The second term
 in Eq.~(\ref{kingpot}), on the other hand,  is not pure gauge. It does lead to
a non-vanishing field strength. It is introduced
in order to cancel   the $x$ dependence of the quark fields  
far from the wall (in the
quark vacua at $z\to\pm\infty$) emerging through the  $x$ dependence of $\sigma (x_n)$,
see  Eqs.~(\ref{wq1}) and (\ref{wq2}). 

\vspace{0.1cm}

To ensure this cancellation we
impose the following boundary conditions for the function $\chi (z)$
\begin{eqnarray}
\label{bchi}
\chi(z)\to\mp 1,\; z\to\pm \infty.
\end{eqnarray}
 This parallels the procedure
outlined  in the toy model of Sect. \ref{toymodel}.
Next, substituting (\ref{wq1}), (\ref{wq2}) and (\ref{kingpot})
in the  action (\ref{redqed}) we arrive at
\begin{eqnarray}
S_{2+1}^{\sigma} &=&\left[\int d^3 x \;\frac12(\pt_n \sigma )^2\right]
\nonumber\\[4mm]
&\times& \int dz\, 
\left\{\frac1{g^2}(\pz\chi)^2 + (1-\chi)^2|\vp^1|^2 +(1+\chi)^2|\vp^2|^2\right\}.
\label{kinsigint}
\end{eqnarray}
The expression in the second line is an ``action" for the $\chi$ profile function.
To get the classical solution for the BPS wall {\em and} the wall world volume
the theory of moduli   we will have to minimize this ``action."

The last two terms in the braces --- the  potential terms in the action for $\chi$
---  come from the 
kinetic terms for the quark fields. The first term  $(\pz\chi)^2$ in the braces ---
the  kinetic term in the action for $\chi$ --- comes from the kinetic term 
of the gauge field. Indeed, the second term in 
Eq.~(\ref{kingpot})   produces
a field strength,
\beq
\label{fmunu}
F_{zn}=\pz \chi \, \pt_{n}\sigma\, .
\eeq
This field strength gives rise to the kinetic term for $\chi$ in
 the action (\ref{kinsigint}).

Now to find the function $\chi$ we have to minimize (\ref{kinsigint}) with
respect to $\chi$. This gives the following equation:
\beq
\label{hieq}
-\pz^2\chi-g^2(1-\chi)|\vp^1|^2+g^2(1+\chi)|\vp^2|^2=0\, .
\label{eqonchi}
\eeq
Note, that the equation for $\chi$ is of the second order. 
This is because the domain wall is no longer BPS state once we
switch on the dependence of the moduli  on the ``longitudinal"
variables $x_n$.

To the leading order  in $\sqrt{\xi}/{\Delta m}$ the solution of
Eq.~(\ref{eqonchi}) can be obtained in the same manner as
we did previously for other profile functions.
Let us first discuss what happens outside the inner part
of the wall. Say, at $z-z_0 \gg R/2$ the profile $|\vp^1|$ vanishes
while $|\vp^2|$ is exponentially close to 
$\sqrt\xi$ and, hence,
\beq
\chi \to - 1 +{\rm const}\, e^{-m_\gamma (z-z_0)}\,.
\label{bvfchi}
\eeq 
The picture at $z_0 -z \gg R/2$ is symmetric, with the
interchange $\vp^1 \leftrightarrow \vp^2$.
Thus,
outside the inner part of the  wall, at $|z-z_0|\gg R/2$, the function $\chi$ approaches its
boundary values $\pm 1$ with the exponential rate of approach.

Of most interest, however, the inside part, 
 the middle domain $M$ (see fig. \ref{syfigthree}).
Here both quark profile functions vanish, and Eq.~(\ref{eqonchi})
degenerates into $\pz^2 \chi =0$. As a result, the solution takes
the form
\beq
\label{hi}
\chi=-2 \frac{z-z_0}{R}\, .
\label{sfchiinm}
\eeq
In  narrow edge domains $E_{1,2}$ the exact $\chi$ 
profile smoothly interpolates between the boundary
values, see Eq. (\ref{bvfchi}),  and the linear  behavior (\ref{hi}) inside the wall. These
edge domains give small corrections to the leading term   in the 
action.

Substituting the solution (\ref{sfchiinm})  in  the $\chi$
action, the second line in Eq.~ (\ref{kinsigint}),  we finally arrive at
\beq
\label{kinsig}
S_{2+1}^{\sigma}=\frac{\xi}{\Delta m}\, \int d^3 x \;\frac12 \, (\pt_n \sigma )^2\, .
\eeq

\vspace{4mm}

  As has been already mentioned previously,  the compact scalar field $\sigma (t,x,y)$
can be reinterpreted to be
 dual to the (2+1)-dimensional Abelian gauge field living on the wall,
see Eq.~(\ref{21gauge}). The emergence of the gauge field on the wall is easy to
understand. The quark fields   almost vanish inside the wall.
Therefore the  U(1)  gauge group is restored inside the wall while it is higgsed
in the bulk. The dual U(1) is in the confinement regime in the bulk.
Hence, the dual U(1) gauge field is localized on the wall, in full accordance with the
general argument of Ref. \cite{DS}.
The compact scalar field $\sigma (x_n)$ living on the wall is a manifestation of this
magnetic localization.

\vspace{1mm}

The result in Eq.~(\ref{kinsig}) implies that the coupling constant of our
effective  U(1)  theory on the wall is given by
\beq
\label{21coupling}
e^2_{2+1}=4\pi^2\,  \frac{\xi}{\Delta m}\, .
\eeq
In particular, the definition of the (2+1)-dimensional gauge field 
(\ref{21gauge}) takes the form
\beq
\label{21gaugenorm}
F^{(2+1)}_{nm}=\frac{e^2_{2+1}}{2\pi} \,\varepsilon_{nmk}\, \partial^k \sigma\, .
\eeq
This finally 
 leads us   to the following
effective low-energy theory of the moduli fields on the wall:
\beq
\label{21theory}
S_{2+1}=\int d^3 x \, \left\{\frac{T_w}{2}\,\,  (\pt_n \zeta )^2+
\frac{1}{4\, e^2_{2+1}}\, (F_{nm}^{(2+1)})^2 +\mbox{fermion terms} \right\}.
\eeq
The fermion fields living on the wall are associated with the four
fermion  moduli
$\eta^{\alpha f}$. On the grounds of (2+1)-dimensional Lorentz symmetry on the wall
we may be certain that these four fermion moduli fields form two (two-component)
Majorana spinors. Thus, the field content of the world sheet theory
we have obtained 
is in full accord with the representation of (2+1)-dimensional extended
supersymmetry (i.e. that with four supercharges\,\footnote{Minimal supersymmetric
theories in 2+1 dimensions have 2 supercharges.}).

Now let us address the question as to the relative
magnitude of the (2+1)-dimensional gauge coupling.
We expect  
that the lightest massive excitations on the wall have mass of the order of the inverse
wall thickness $\sim 1/R$. 
Since
\beq
\frac{1}{R}=\frac{g^2}{4}\, \frac{\xi}{\Delta m} = e^2_{2+1}\, \frac{g^2}{16\pi^2}\,,
\label{rel21cc}
\eeq
 our (2+1)-dimensional
coupling constant (\ref{21coupling}) is large
in this scale, 
so that 
the theory on the wall is in the strong coupling regime. This could 
have been expected since 
it is a  U(1)  theory of the
dual degrees of freedom (magnetic charges),
 as  will be discussed in more detail  in Sect. \ref{stringend}.

\subsection{Nonzero modes}
\label{nonze}

It is not difficult to see
that the lightest massive excitations are associated
with perturbation of the size of the middle domain $M$. When its
  thickness ``breathes," this gives rise to  the softest mode.
The mass of  the softest mode can be readily estimated to be of the order of
$m_\gamma^2 /\Delta m \sim R^{-1}$.

\subsection{\ntwo supersymmetry and the multiplicity
of the domain walls}
\label{ntwomultipli}

We have just demonstrated that one can find 1/2 BPS saturated domain wall
in the \ntwo SQED with the generalized Fayet-Iliopoulos term.
This fact {\em per se} has far reaching consequences for the
world-volume theory as well as for the multiplicity of the
domain walls. Indeed, 2+1 dimensional world-volume theory
must have {\em four} supercharges.  This implies with necessity
that there are {\em two} (massless) boson fields in the 
world-volume theory.  On general grounds they can form either
a chiral supermultiplet   of \ntwo in 1+2 or a vector supermultiplet 
of a world-volume theory with U(1) gauge invariance (linear supermultiplet). 
As long as the fields are massless, there is a duality between these two descriptions.

Supersymmetrization  requires two (Majorana) two-component fields  in
the world-volume theory. The dimension  of the supermultiplet we deal with is four. 
The question to be discussed is: how many distinct domain
walls we have?  This question is meaningful in  light
of the recent finding \cite{KSS,AVafa,Ritz:2002fm}
of nontrivial multiplicity
of the domain walls in \none super-Yang-Mills theories.

In addressing this question we have to explain our convention.
Every (2+1)-dimensional domain wall emerging in the
(3+1)-dimensional theory has a translational and supertranslational moduli.
If we quantize these moduli in a finite volume
the corresponding wall multiplicity is two. This part of the moduli dynamics is trivial,
however, and can (and should) be factored out. 
That's what we will always do.
When we speak of the wall multiplicity
we discard the above trivial degeneracy and focus exclusively
on possible extra degeneracy not associated with the
(super)translational moduli. 
With this convention one can demonstrate \cite{KSS,AVafa,Ritz:2002fm} that
the number of   distinct domain walls
in SU($N$) super-Yang-Mills theories is in fact $N$.

Following this convention
we can say that   the 1/2 BPS wall
we have constructed  has multiplicity two. This is rather obvious by itself,
since in the case at hand the  original (3+1)-dimensional theory we began with
had eight
supercharges. What is remarkable is that
the above statement will hold even if we break \ntwo
of the original theory down to \none.

To see that the wall multiplicity is two it
is sufficient to compactify the
longitudinal directions $x$ and $y$.
Then the reduced moduli field theory on the wall\,\footnote{By reduced we mean
that the translational modulus and its superpartner are factored out.}
becomes quantum mechanics of one
real variable $\sigma (t)$ defined on a circle, 
$$
\sigma + 2\pi \leftrightarrow \sigma\,,
$$
and two fermion variables $\psi ( t)$ and $\bar\psi (t)$. 
This supersymmetric quantum mechanics has the ground state at zero energy
which is doubly degenerate. This double degeneracy is protected against nonsingular
perturbations, such as generation of the potential term for $\sigma$,
which, generally speaking,
 might occur if \ntwo is broken down to \none.
The potential term is {\em not} generated for our domain wall;
presumably, it is generated in the case considered in Ref. \cite{KS}.

\section{The ANO strings}
\label{anostrings}
\setcounter{equation}{0}

In string theory 
gauge fields are localized  on   D branes  
because fundamental open strings can end on a D brane. 
Our task now is to investigate to which extend
 this picture --- a flux tube ending on 
the critical
 domain wall --- holds in  field theory. We will see that the answer to this question is positive:
our 1/2 BPS domain wall does allow for the magnetic
flux tubes to end on it. 

As we have already explained, both quark vacua in our \ntwo QCD give rise to
confinement phase for monopoles \cite{HSZ,Ymc4}. The monopoles 
themselves are very
heavy in  the quark vacua; the  monopole mass 
is of the order of $m/g^2$. Hence, the monopoles can be
considered as    probes for confinement. Here we 
deal with the Abelian confinement,
due to  the ANO flux tubes which stretch between monopoles 
and antimonopoles.

Now we will demonstrate, through 
an exact solution, that the ANO string can end   on the domain wall
interpolating between two quark vacua. Imagine a monopole placed at
some point in the bulk far away from the domain wall. The 
magnetic flux of the monopole
is trapped inside the flux tube in the bulk. When the
tube joins the wall   
the magnetic flux of the tube becomes electric flux of the dual U(1) theory on the
wall; therefore it spreads out along the wall. The end point of the tube
on the wall   plays the role of an electric charge in the (2+1)-dimensional 
U(1) gauge theory. Ending the flux tube on the wall, rather 
than letting the tube go through,
is energetically advantageous. The field configuration with
a string attached to the wall is 1/4 BPS saturated in our model. 
In other words, the theory localized on
the string-wall junction has two supercharges.

Before delving in details of the string-wall junction
 construction
(which will be done shortly)
we have to briefly review 
 the critical  ANO strings in the Seiberg-Witten theory
\cite{FG,Ymc4,EFMG,VY,KoS}. Our presentation follows 
 that of Ref.~\cite{VY}, the only difference is that Ref. \cite{VY} deals
with the critical strings in the monopole/dyon vacua 
while here we are interested in the quark 
vacua.

Let us consider, say, the ANO strings in the vacuum (\ref{a1}). The field $a$
is irrelevant for the string solution so we can put it equal to its VEV
(\ref{a1}) and drop from the effective QED 
Lagrangian, Eq.~(\ref{qed}). In the vacuum (\ref{a1})
 the second
quark flavor $q^2$ is heavier than the first one and we can ignore it either.
Moreover, it turns out that  the string solution we are after admits   
the same {\em ansatz} (\ref{qqt}) we exploited for finding the wall solution.

With all these simplifications the effective action of our model (\ref{redqed})
becomes
\beq
\label{ahbps}
S_{\rm str}=\int d^4x\left\{\frac1{4g^2}F^2_{\mu\nu}
 +|\nabla_\mu\varphi^1|^2+\frac{g^2}{8}\left(
|\varphi^1|^2-\xi\right)^2\right\}.
\eeq
First, let us compare this theory with the   general  
 Abelian Higgs model in which the ANO vortices are known to exist 
\cite{ANO}. The action of the Abelian Higgs model reads
\beq
\label{ah}
S_{\rm AH}=\int d^4x\left\{\frac1{4g^2}F^2_{\mu\nu}+
|\nabla_\mu\varphi^1|^2
+\lambda \left(|\varphi^1|^2-\xi\right)^2\right\}.
\eeq
We see that the model (\ref{ahbps}) which appears in \ntwo QED with
the  FI
term (the same model appears in \none SUSY with  the FI term
\cite{DDT}) 
corresponds to a special value of the coupling
$\lambda$,
\beq
\label{lambda}
\lambda\ =\ \frac{g^2}8\ .
\eeq
In the model (\ref{ah}) photon has mass (\ref{mgamma}) while quark
$\varphi^1$ acquires the mass
\beq
\label{higgsm}
m^2_H\ =\ 4\lambda\xi\ .
\eeq

\vspace{0.1cm}

For generic values of $\lambda$ in Eq.~(\ref{ah}) the quark mass (the inverse correlation
length) and the photon mass (the inverse penetration depth) are distinct. 
Their ratio is an important
parameter characterizing the type of the superconductor
under consideration. Namely,
for $m_H<m_\gamma$ one deals with the type I superconductor in
which two strings at large separations attract each other.
On the other hand, for 
$m_H>m_\gamma$ the 
superconductor is of type II,  in which case two strings
at large separation repel each other. This behavior is related to the fact
that the scalar field generates  attraction between two  vortices,
while the electromagnetic field generates  repulsion.

Now we see that with the choice (\ref{lambda}) for $\lambda$
the masses are equal, 
\beq
\label{bpscond}
m_\gamma\ =\ m_H\,,
\eeq
and the superconductor is exactly at the border between type I and type II.
The relation (\ref{bpscond}) has important consequences. It means
that the ANO string satisfies the first order equations and
saturates the Bogomolny bound \cite{B}. The BPS 
strings do not interact.

To get the BPS equations for the string in the model (\ref{ahbps}) 
(as well as the BPS bound for its tension)
it is convenient to perform the Bogomolny completion
of the action,  
\begin{eqnarray}
\label{bogst}
 T_{\rm str}\ &=& \int d^2 x\left\{\frac1{2g^2}\left[F^{*}_{3}-
\frac{g^2}{2}\left(|\varphi^1|^2-\xi\right)\right]^2\right. +
\nonumber\\[3mm]
& + & 
\left.\left|(\nabla_1-i
\nabla_2)\varphi^1\right|^2\right\} +2\pi\xi n\, .
\end{eqnarray}
Here we assume that the string is aligned along the $x_3=z$ axis
with its center at the point $x_1=x_2=0$ and, moreover,
\beq
F^{*}_n= \frac{1}{2}\, \ve_{nmk}\, F_{mk}\,,\qquad n,m,k=1,2,3\,.
\label{stardef}
\eeq
 The last term in (\ref{bogst}) measures the quantized magnetic
flux of the vortex, $n$ is the winding number. 
For simplicity we will consider the minimal winding, $n=1$.

\vspace{1mm}

As usual, implications of the Bogomolny completion are immediate.
The energy minimum   is reached if the first two terms in the braces
vanish individually.
Then the string tension is, obviously,
\beq
\label{sten}
 T_{\rm str}\ =\ 2\pi\xi \, .
\eeq
The 
vanishing of the first two terms in the braces
imply the following 
first order equations:
\begin{eqnarray}
&& F^{*}_{3}-\frac{g^2}{2}\left(|\varphi^1|^2-\xi\right)=0\,,
\nonumber\\[3mm]
&& (\nabla_1-i\nabla_2)\varphi^1=0\, .
\label{sfoe}
\end{eqnarray}
Certainly, these equations are well studied in the literature \cite{B}.

\vspace{1mm}

The classical ANO vortex solution for the fields $\vp^1$ and
$A_\mu$ is obtained in  the standard {\em ansatz}, 
\begin{eqnarray}
\label{stans}
\varphi^1(x) &=& \phi(r)e^{-i\alpha}\ ,\nonumber\\[3mm]
A_i(x) &=& 2\,  \varepsilon_{ij}\frac{x_j}{x^2}\ [1-f(r)]\ ,
\end{eqnarray}
where $i,j=1,2$, $r=\sqrt{x^2_i}$ and $\alpha$ is the polar angle in the 
(1,2) plane (Fig. \ref{syfigtwo}). 
The real profile functions $\phi(r)$ and $f(r)$ 
satisfy the boundary conditions 
\begin{eqnarray} 
\label{sbc}
&& \phi(0)=0\ , 
\quad f(0)=1\ , \nonumber\\[2mm]
 && \phi(\infty)=\sqrt{\xi}\ , \quad 
f(\infty)=0\ , 
\end{eqnarray}
which ensure  that the scalar field reaches its VEV $\sqrt{\xi}$ at 
infinity and the vortex carries precisely one unit of the magnetic flux.

With this {\em ansatz}, the first order equations (\ref{sfoe}) become
\begin{eqnarray}
\label{ansfoe}
&&\phi'-\frac1r\ f\phi\ =\ 0\ , \nonumber\\[2mm]
&& -\frac1r\ f'+\frac{g^2}{4}(\phi^2-\xi)\ =\ 0\ .
\end{eqnarray}
These equations, together  with the boundary conditions (\ref{sbc}), can be solved
numerically.

In supersymmetric theories the Bogomolny bound (\ref{sten}) for 
the string tension 
can be viewed as a central charge (in the anticommutator $\{
Q_\alpha \,, \bar Q_{\dot\beta}\}$) of the supersymmetry algebra.
The first order equations (\ref{sfoe}) can be
obtained by requiring    half of the supercharges to act trivially on the 
string solution \cite{HS,DDT,GS,VY}. In order to see which 
particular supercharges act
trivially we write down all relevant SUSY transformations (\ref{str}) in our effective
QED and put the fermion components to zero. Dropping the fields $a$ and $q^2$
(as was discussed in the beginning of this section)
and assuming  that all fields depend only on 
the coordinates
$x_1$,  $x_2$ we arrive at the same equations 
(\ref{sfoe}) provided that the SUSY transformation parameters $\ve^{\alpha f}$
and $\bar{\ve}_{\dot{\alpha}}^{f}$
are subject to the following constraints:
\begin{eqnarray}
&& \ve^{12} =- \ve^{11}\, , \quad \bar{\ve}_{\dot{1}}^{2} =- \bar{\ve}_{\dot{1}}^{1}   \,,  
    \nonumber\\[3mm]
&& \ve^{21} =\ve^{22}\, , \qquad \bar{\ve}_{\dot{2}}^{1} =\bar{\ve}_{\dot{2}}^{2}\, .
\label{sepsilon}
\end{eqnarray}
These conditions select those supercharges which act trivially on the BPS string
solution. Moreover, they explicitly   show  that our ANO string is 1/2
BPS-saturated. Note, however, that the conditions (\ref{sepsilon}) are different
from those in Eq.~(\ref{wepsilon}), so that four supercharges preserved by the string and
four supercharges preserved by the wall are not the same.

\section{String ending on the wall}
\label{stringend}
\setcounter{equation}{0}

In this section we 
derive BPS equations and find a 
1/4 BPS
solution for the wall-string junction.
We analyze qualitative features of the solution
and investigate how the magnetic flux of the string
gets spread inside the wall. 

\subsection{First order equations for a string
ending on the wall}
\label{stringending}

It is natural to assume that at large distances from the string end
point  at $r=0$, $z=0$,
the wall is almost parallel to the $(x_1,x_2)$ plane while the string
is stretched along the $z$ axis, see Fig. \ref{syfigtwo}. Since 
both solutions, for the string and the wall,  were obtained using the 
{\em ansatz} (\ref{qqt}) we restrict our search for  the wall-string junction
to the same {\em ansatz}. As usual, we look for a static solution
assuming that  all relevant fields can depend only on $x_n$, ($n=1,2,3$).

First, we have to decide which particular combinations of supercharges
act trivially on the wall-string junction configuration.
To this end we impose both, the wall conditions (\ref{wepsilon}) and the string
conditions (\ref{sepsilon}), simultaneously. It turns out that there is
a nontrivial solution with
\begin{eqnarray}
\label{epsilon}
&& \ve^{12} =- \ve^{11}\ ,  \nonumber\\[2mm]
&& \ve^{21} =\ve^{22},
\end{eqnarray}
while the  parameters $\bar{\ve}$ are given in Eq.~(\ref{wepsilon}).
We see that  all eight SUSY parameters can be expressed in terms of two
arbitrary parameters  $\ve^{11}$ and $\ve^{22}$. Thus, the 
string-ending-on-the-wall configuration is 1/4 BPS saturated.

Now we substitute Eq.~(\ref{epsilon}) in the SUSY transformations
(\ref{str}) and put the fermion components to zero.
This leads us to the following first order equations:
\beqn
&& F^{*}_1-iF^{*}_2 - \sqrt{2}(\pt_1-i\pt_2)a=0\, ,
\nonumber\\[3mm]
&& F^{*}_{3}-\frac{g^2}{2} \left(\left| \varphi^{A}\right|^2-\xi\right) 
-\sqrt{2}\, \pt_3 a =0\, ,
\nonumber\\[3mm]
&& \nabla_3 \vp^A =-\frac1{\sqrt{2}}\vp^A(a+\sqrt{2}m_A)\, ,
\nonumber\\[4mm]
&& (\nabla_1-i\nabla_2)\varphi^A=0\, .
\label{foe}
\eeqn
These equations   generalize the first order equations for the wall
and for the string ($F^*$ is defined in Eq. (\ref{stardef})). 

It is instructive to  check that both, the wall and the string 
solutions, separately,
satisfy these equations. Start from the wall. In this case the
gauge field is pure gauge (see Eq.~(\ref{wgpot})),
and all fields depend only on $z$. Thus, the 
first and the last equations in (\ref{foe}) are trivially satisfied.
The component of the gauge field $F^{*}_3$ vanishes in the second equation
and this equation reduces to the second equation in Eq.~(\ref{wfoe}). The 
 third equation in (\ref{foe}) coincides  with  the first one
in (\ref{wfoe}).

For the string which lies, say, in the vacuum (\ref{a1}), the second quark flavor
vanishes,
 $q^2=0$, while $a$ is given by its VEV.
The electromagnetic flux is directed along the $z$ axis, so 
that $ F^{*}_1=F^{*}_2=0$.
All fields depend only on the  coordinates $x_1$ and $x_2$. Then the first and 
the third equations in (\ref{foe}) are trivially satisfied. The second equation
reduces to the first one in Eq.~(\ref{sfoe}). The last equation in (\ref{foe})
for $A=1$ reduces to the second equation in (\ref{sfoe}), while for $A=2$
this equation is trivially satisfied, {\em quod erat demonstrandum}.

\subsection{The string-wall junction (Solution for a string ending on the wall)}
\label{stringendingonwall}

Needless to say that the solution of first order equations (\ref{foe}) for a string
ending on the wall can be found only numerically especially near the end point of
the string  where both the string and the wall profiles are heavily deformed.
However, far away from the end point of the string, deformations are
weak and we can find the asymptotic behavior analytically.

Let the string be on the $z>0$ side of the wall, where the vacuum 
is given by Eq.~(\ref{a2}), see Fig. \ref{syfigtwo}. 
Consider first the region $z\to \infty$ far away
from the string end point at  $z\sim 0$. Then the solution to
(\ref{foe}) is given by an almost unperturbed string.
Namely, at $z\to\infty$ there is no $z$ dependence
to the leading order, and, hence,    the following {\em ansatz}
  for fields $A_{1,2}$ and $\vp^2$ is appropriate:
\begin{eqnarray}
\label{singans}
\varphi^2(x) &=& \phi(r)\ ,\nonumber\\[3mm]
A_i(x) &=& -\varepsilon_{ij}\frac{x_j}{x^2}\ f(r)\,.
\end{eqnarray}
It differs from the one
in Eq.~(\ref{stans}) by a gauge transformation, to the ``singular''  gauge,
in which the scalar field $\vp^2$  is aligned along its 
VEV at $r\to\infty$, $z\to\infty$.
The profile functions $\phi(r)$ and $f(r)$ satisfy the boundary conditions
(\ref{sbc}).
We also take the fields  $A_{3}$, $A_{0}$ and  $\vp^1$
to be  zero, with  $a$ equal  to  its VEV (\ref{a2}). Then Eqs.~(\ref{foe})
reduce to those of   (\ref{ansfoe}). The latter  have a standard solution of
the unperturbed ANO string. On the other side of the wall, at $z\to -\infty$,
we have an almost unperturbed first vacuum with the fields  given by their
VEV's (\ref{a1}) and (\ref{q1}).

\begin{figure}[h]  
\epsfxsize=7cm
\centerline{\epsfbox{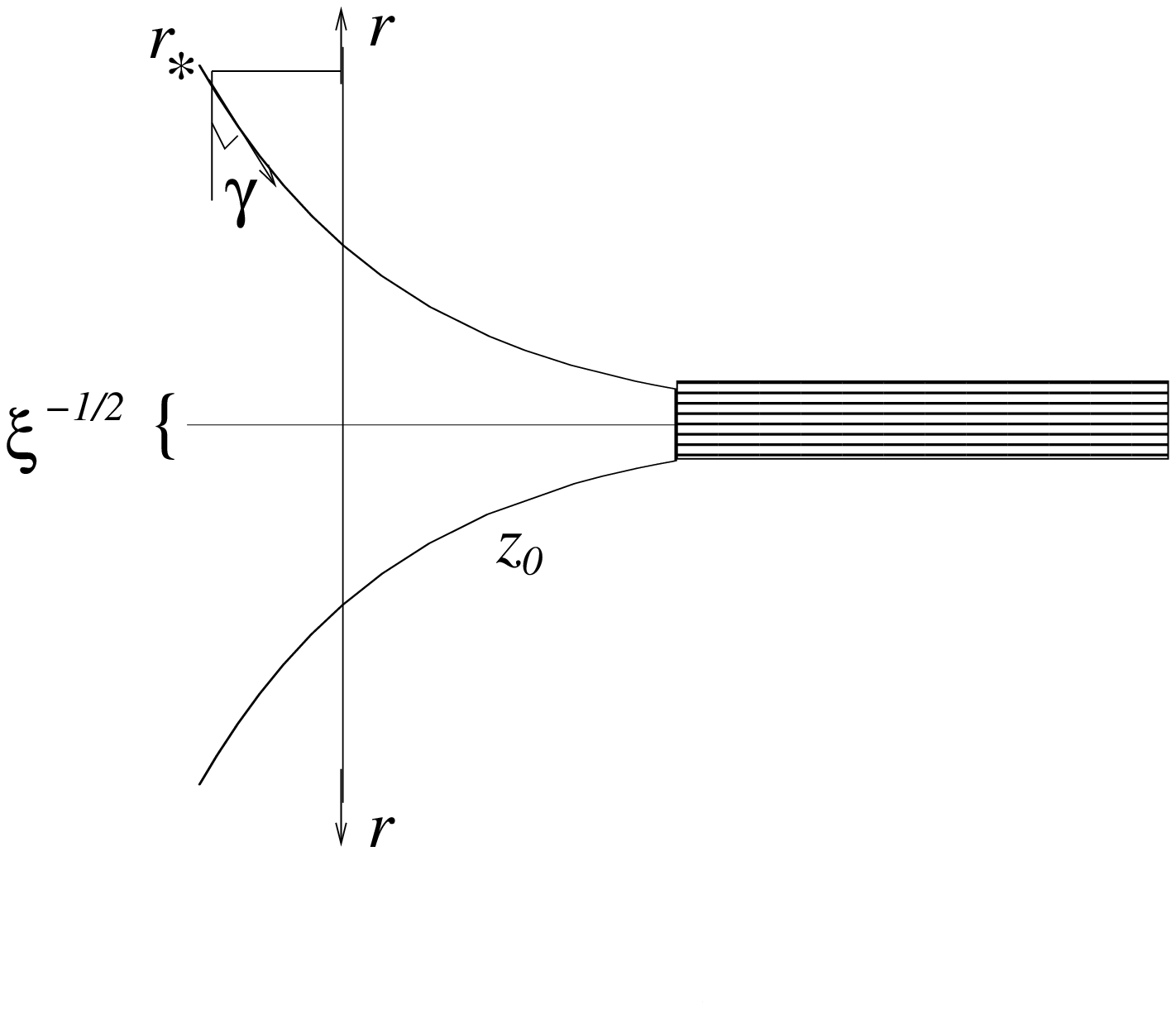}}
 \caption{Bending of the wall due to the string-wall junction.
The flux tube extends to the right infinity. The wall
profile is logarithmic at transverse distances larger than
$\xi^{-1/2}$ from the string axis. At smaller distances 
the adiabatic approximation fails.}
\label{syfigfive} 
\end{figure}

Now consider the domain $r\to\infty$ at small $z$. In this domain the solution
to (\ref{foe}) is given by a  perturbation of the wall solution. Let us
use the ansatz in which the solutions for the fields $a$ and $q^{A}$ are given by
the same equations (\ref{a}), (\ref{wq1}) and (\ref{wq2})  in which the 
size of the wall is still given by  (\ref{R}), and {\em the only modification} is
that the position of the wall $z_0$ and the phase $\sigma$  now become
slowly-varying functions of $r$ and $\alpha$ (i.e. the polar coordinates on the $(x_1,x_2)$ plane). It is quite obvious that $z_0$ will depend only on $r$,
as schematically depicted in Fig. \ref{syfigfive}.
The physical meaning of this ``adiabatic"
approximation is as follows:
the massive excitations of the wall, responsible for its structure, 
are assumed to be absent; we study the impact of the string-wall 
junctions on the massless moduli.

As long as the second and the third equations in (\ref{foe}) do not contain
derivatives with respect to $x_i$, $i=1,2$, they are satisfied identically for any
functions $z_0(r,\alpha)$ and $\sigma (r,\alpha)$ (note, that $F^{*}_3=0$,
the field strength is parallel to  the  domain wall plane
and  $A_z= \sigma(r,\alpha)\pz \beta(z)$, see 
Eq.~(\ref{wgpot})).

However, the  first and the last equations in (\ref{foe}) become nontrivial.
Consider the first one. Inside the string the electromagnetic field is
directed along the $z$ axis and its flux is given by $4\pi$. This flux
 is spread out  inside the wall and 
directed  almost along $x_i$ in the $(x_1,x_2)$ plane at large $r$.
Since the flux  is conserved,  we have
\beq
\label{flux}
F^{*}_{i}=\frac{2}{R}\,\, \frac{x_{i}}{r^2}
\eeq
inside the wall at $|z-z_0(r,\alpha)|<R/2$.

Substituting this in the first equation in (\ref{foe}) and assuming
that $z_0$ depends only on $r$ we then get
\beq
\label{derz}
\pt_{r}z_0=-\frac1{\Delta m r}\, .
\eeq
Needless to say that our adiabatic approximation holds
only provided the above derivative is small, i.e.
sufficiently far from the string,
 $\sqrt{\xi} r\gg1$.

The solution to this equation is straightforward,
\beq
\label{z0}
z_0=-\frac1{\Delta m }\ln {r} + {\rm const}.
\eeq
We see that the wall is logarithmically bent according to the Coulomb law 
in 2+1 dimensions\footnote{The logarithmic bending in two spatial dimensions
was used by Witten to explain logarithmic running of coupling constant
of \ntwo theories within the brane approach \cite{W54}.} (see Fig. \ref{syfigfive}). 
 This bending produces a balance of forces
between the string and the wall in the $z$ direction so that the whole configuration is
static. To see that this is indeed the case,
please, observe   that the force of the string is equal to the 
string tension $2\pi\xi$ (see Eq.~(\ref{sten})). On the 
other hand,  the force of the wall in
$z$-direction at some point $z_{*}$ is given by the wall tension $\xi \Delta m$
times the length of the circle $2\pi r_{*}$ ($r_{*}$ corresponds to 
$z_{*}$ via Eq.~(\ref{z0})) times the angle $\gamma$ following\,\footnote{One should remember
that $r_*\gg (\Delta m)^{-1}$, so that $\gamma \approx \tan\gamma\approx\sin \gamma $.}
 from Eq.~(\ref{derz}), which  projects 
the force of the bent wall onto the $z$ axis. This gives $2\pi\xi$, which  
precisely  coincides with
 the string tension.

Now let us consider the last equation in (\ref{foe}). First, let us work out the gauge 
potential which enters the covariant derivatives in this equation. In order to
produce the field strength (\ref{flux}) $A_\mu$  should reduce to
\beqn
A_{i} &=& \frac{2}{R}\,\ve_{ij}\frac{x_{j}}{r^2}[z-z_0(r)]
+ \beta(z)\pt_i \sigma(x_j)\,, \qquad i=1,2\, , 
\nonumber\\[4mm]
 A_{0} &=& 0,\qquad A_z=\sigma(x_i)\pz \beta(z)\, ,
\label{gaugepot}
\eeqn
where we also include  pure gauge terms, see Eq.~(\ref{wgpot}).
Consider first the region near the edge of the wall at $z-z_0 \sim -R/2$.
Near this edge the first quark field $\vp^1$ is not zero. Substituting 
(\ref{wq1}) in  the last equation in (\ref{foe}) and using (\ref{gaugepot})
and (\ref{derz}) we get
\beq
\label{dersig}
\frac{\pt \sigma}{\pt \alpha}=1,\qquad \frac{\pt \sigma}{\pt r}=0\, .
\eeq
The solution to these equations is
\beq
\label{sigmaalpha}
\sigma =\alpha\, .
\eeq
This vortex solution is certainly expected and welcome. In terms of 
the dual gauge field localized on the wall, this solution reflects
nothing but the unit source charge.
(If we   consider the other edge of the wall near $z-z_0 \sim R/2$
and substitute (\ref{wq2}) into the last equation in (\ref{foe})
we get the same equations (\ref{dersig}) for $\sigma(r,\alpha)$.)

The above relation between the vortex solution and the unit source charge
requires a comment.
One can identify the 
compact scalar field $\sigma$ with the electric  field living
on the domain wall world volume via (\ref{21gaugenorm}). Then
result (\ref{sigmaalpha}) gives (\ref{elf}) for this electric
field. In the proper normalization we  have
\beq
\label{elfnorm}
F_{0i}^{(2+1)}=\frac{e^2_{2+1}}{2\pi}
 \,\,\frac{x_i}{r^2}\,,
\eeq
where the (2+1)-dimensional coupling is given by (\ref{21coupling}).

 This  is a field of a point-like electric charge 
in 2+1 dimensions placed at $x_i=0$.
The interpretation of this result is that the string end point   
 on the wall plays a role of the electric charge in 
the  dual U(1) theory on the wall.

Our string-wall junction solution 
explicitly demonstrates another rather apparent aspect
of the problem at hand. Indeed, when the string ends on the wall,
and the magnetic flux it brings with it spreads out inside the wall,
the overall energy of the configuration is minimized.
Indeed, the energy of the flux tube grows linearly with its dimension,
while when the flux is spread inside the wall, the energy 
grows only logarithmically with the dimension of the domain over which the flux is spread. It is certainly no accident that the wall bending is
logarithmic.

\section{Flow to \none theory}
\label{flow}
\setcounter{equation}{0}

It is high time now to address
the question  what happens with the moduli theory once 
 we include subleading in $\mu$
terms that break \ntwo supersymmetry
down to \none.  At the level of the low-energy QED (\ref{qed}) 
to which we limit ourselves in the present paper this amounts to
taking into account the $a$ dependence of the function $f(a)$ in Eq.~(\ref{qed}),
\beq
\label{fofa}
f(a)\equiv -2\sqrt{2}\, \mu\,  \frac{\partial u}{\partial a}=- 2\sqrt{2}\, \mu \, a\,,
\eeq
while previously $f(a)$ was set to be constant, $f(a) = \xi$, in which limit
we deal with the fully blown \ntwo.

Let us first discuss the value of an \ntwo breaking parameter on general grounds.
It is quite clear that in both vacua
this parameter can be estimated as
\beq
\frac{\mu\, \delta a_{\rm char}}{\xi}
\sim \frac{\mu\, \sqrt\xi}{\xi}\sim
\sqrt{\frac{\mu}{m}}\,.
\label{en2sb}
\eeq
This estimate is confirmed, in particular,
 by 
the calculation of the fermion mass shift presented in
Sect. \ref{fmnbroken}. However when one deals with 
collective excitations of the
domain wall (other than the zero modes),
then the \ntwo breaking parameter is larger, since in this case
$\delta a_{\rm char}$ is of the order of $\Delta m$.
Then, the \ntwo breaking parameter
is of the order of
\beq
\label{n2brw}
\frac{\mu\Delta m}{\xi}.
\eeq
Although
this is still a small parameter   it is nevertheless
larger than the one in (\ref{en2sb})
under our choice
(\ref{mxi}). This parameter controls the splitting of massive \ntwo multiplets in
the world-volume theory on the wall. 
It is not difficult to calculate this mass splitting, and so we did.
If we substitute the expression for $\xi$ from
Eq.~(\ref{xim}) here, then (\ref{n2brw}) reduces to $\Delta m/m$.
Note, however, that we can ignore the relation (\ref{xim}) coming from
underlying non-Abelian theory and consider the $U(1)$ theory (\ref{qed})
on its own right. Then the \ntwo limit is characterized by parameters
$\xi$ and $m^A$ while $\mu$ controls \ntwo breaking down to \none.
In this setup the parameter of \ntwo breaking on the wall
is given by (\ref{n2brw}) and goes to zero at $\mu\to 0$ while $\xi$
is fixed.

\vspace{2mm}

It is easy to see that with $ f(a)\neq$ const the Bogomolny completion
for the domain wall configuration is still possible, 
  and we end up with the following first order equations:
\begin{eqnarray}
\nabla_z \vp^A &=& -\frac1{\sqrt{2}}\vp^A\left(a+\sqrt{2}m_A\right)\, ,
\nonumber\\[3mm]
\label{n1wfoe}
\pz a &=&- \frac{g^2}{2\sqrt{2}}\left(|\vp^A|^2-f(a)\right)\, ,
\end{eqnarray}
where  the same {\em ansatz} as in Eq.~(\ref{qqt}) is used. 
These equations
are quite similar to the  equations (\ref{wfoe}) for \ntwo theory, 
the only difference being  that
the constant FI parameter $\xi$ is now replaced by a
linear (and slowly-varying)
function $f(a)$.
The tension of this domain wall is still given by Eq.~(\ref{wten}).

The emergence of the first order equations means that the wall
at hand 
is still   BPS saturated (albeit in \none theory).
 This can be most straightforwardly seen by
 noting that
a half of the relations (\ref{wepsilon}) survive breaking of \ntwo
to \none. 
 Now we have four parameters $\ve^{\alpha 1}$
and $\bar{\ve}_{1\dot{\alpha} }$, subject to 
two constrains,
\beq
\label{n1epsilon}
\bar{\ve}_{1\dot{2} }=-i\ve^{2 1}\,,\qquad
\bar{\ve}_{1\dot{1} }=i\ve^{1 1}\, .
\eeq
This leaves us with two supercharges which act trivially on the wall
solution.

At small $\mu$ and $\Delta m\ll m$ the solution to first order equations
(\ref{n1wfoe}) is 
a small perturbation of the domain wall solution presented in  Sect. \ref{wallsolution}.
Thus, it is still parametrized by two collective coordinates (moduli),
$z_0$ and $\sigma$. In the exact \ntwo limit
the moduli fields $\zeta (t,x,y) $ and $\sigma (t,x,y)$  formed the bosonic part of \ntwo vector multiplet
in 2+1 dimension. 
 Now the question is: do these fields
split once we break \ntwo supersymmetry softly down to \none?

Naively one might think that 
while the ``translational" field $\zeta (t,x,y) $ 
certainly remains massless, the second modulus field $\sigma (t,x,y)$
becomes  massive   acquiring a small mass, so that the \ntwo
supermultiplet is split in two \none
supermultiplets. A particular mechanism for such splitting
was suggested in Ref.~\cite{AVafa} for the case of the  domain wall 
interpolating between the monopole and dyon vacua. Namely, it was suggested
that a
  Chern-Simons term is generated in the (2+1)-dimensional gauge theory
on the wall, generating a
 mass to the  U(1)  gauge field.
Although this conjecture is very relevant and will be exploited in
Sec. \ref{mult-ksy}, we will prove momentarily
that  the splitting does
{\em not} take place for the wall interpolating between the quark vacua ---
the object of our study in this paper. The field $\sigma$ stays massless.
In other words, the moduli field theory on the
wall exhibits a supersymmetry enhancement ---
the particle contents is characteristic of the theory with
four supercharges, rather than two supercharges one might expect
to operate on the world sheet of 1/2 BPS domain wall in \none theory.

First, this conclusion 
follows from symmetry arguments. Indeed, the
phase $\sigma$ is associated with the global  U(1)  rotation which is
a part of the flavor  SU$(N_f=2)$ broken down to  U(1)  by 
the mass difference, $\Delta m\neq 0$.
This symmetry is a vector-like  and therefore is not anomalous.
Thus, we have an exact global symmetry broken down on the domain wall solution.
This implies the inevitability
of the zero mode associated with the 
collective coordinate  $\sigma$. With two massless moduli,
$\zeta (t,x,y) $ and $\sigma (t,x,y)$, \none supersymmetry on the
world sheet would imply two massless Majorana fields in the fermion sector.
The fermion moduli are related to the fermion zero modes.
It is instructive to check, by counting the fermion zero modes,
that in the fermion sector we deal with two massless Majorana fields
localized on the domain wall.
We will now show that we have four fermion zero modes on the  wall solution
(two plus two complex conjugated).

To this end we use the Jackiw-Rebbi theorem  which tells us that we have
exactly two normalizable fermion zero modes (one two-component 
real fermion field)
per each eigenvalue of the fermion
mass matrix which changes its sign on the wall solution \cite{jackiw}.
The fermion mass matrix in softly broken SQED is given by
\beqn
S_{\rm mass}^{\rm ferm} &=& \frac1{\sqrt{2}}\int d^4 x \left\{
\bar{q}_{Af}(\lambda^f\psi^{A}) -(\tilde{\psi}_{A}\lambda_f)q^{fA}
\right.
\nonumber\\[3mm]
&-& \left. (a+\sqrt{2}m_{A})(\tilde{\psi}_{A}\psi^A) -\frac{\mu}{\sqrt{2}}(\lambda^2)^2
\right\} .
\label{yukawa}
\eeqn
The last term explicitly breaks \ntwo supersymmetry and $SU(2)_{R}$
global symmetry because it depends only on the $f=2$ components of 
the field $\lambda^f$.

The fermion mass terms and Yukawa couplings in (\ref{yukawa}) give us
a $6\times 6$ fermion mass matrix. To study its eigenvalues we calculate its
determinant. The general expression for the determinant is 
rather complicated; therefore, we study it approximately in different domains
of the wall profile, see Fig.~\ref{syfigthree}. The six eigenvalues of the mass
matrix can be naturally divided in two classes:
 four ``large'' eigenvalues and two ``small'' ones.

\subsection{Fermion modes, \ntwo limit}
\label{fmn2}

Consider first the \ntwo limit when the last term in Eq.~(\ref{yukawa}) is ignored.
Two ``small'' eigenvalues approach the photon mass (\ref{mgamma})
in both quark vacua,
\beq
\label{srhoass}
|\rho^s_{1,2}|\to m_{\gamma},\qquad  z\to\pm\infty.
\eeq
The ``large"  eigenvalues behave as follows. In the left vacuum
two of them approach the photon mass while the other two approach
the larger value,  $\Delta m$,
\beq
\label{lrhoassleft}
|\rho^l_{1,2}|\to m_{\gamma},\qquad |\rho^l_{3,4}|\to \Delta m,\qquad z\to-\infty,
\eeq
while in the right vacuum their role is interchanged, namely
\beq
\label{lrhoassright}
|\rho^l_{3,4}|\to m_{\gamma},\qquad |\rho^l_{1,2}|\to \Delta m, \qquad z\to\infty.
\eeq
This behavior is, of course, in perfect agreement with the mass
spectrum of the theory in both quark vacua found in Sect. \ref{n2sqed}.

In the middle  region $M$ (see  Fig.~\ref{syfigthree}) ``large'' eigenvalues 
interpolate between these two values. They   always remain large inside the wall and
clearly do not cross the zero. ``Small'' eigenvalues are exponentially small
inside the wall and, therefore, one  needs to carry out a
more careful study.
In the middle domain $M$ these eigenvalues are given by
\beq
\label{srho}
\rho^s_{1,2}=\pm\frac12\left[\frac{g^2|\vp^1|^2}{m_1+a/\sqrt{2}} + 
\frac{g^2|\vp^2|^2}{m_2+a/\sqrt{2}}\right],
\eeq
where  the functions $a$ and $\vp^{A}$ are given by the domain wall
solution (\ref{a}),  (\ref{wq1}),  and (\ref{wq2}).
Two ``small'' eigenvalues (\ref{srho}) clearly cross the zero at $z=z_0$,
right in the middle of the domain wall.
This is in accordance with our previous  conclusion that we have four
fermion zero modes in \ntwo limit.

\subsection{Fermion modes, \ntwo broken down to \none }
\label{fmnbroken}

 Now let us take into
account the \ntwo breaking term in Eq.~(\ref{yukawa}). Then
the first eigenvalue in (\ref{srho}) gets  an additional contribution
\beq
\label{srho1}
\rho^s_{1}=\frac12\left[\frac{g^2|\vp^1|^2}{m_1+a/\sqrt{2}} + 
\frac{g^2|\vp^2|^2}{m_2+a/\sqrt{2}}+2g^2\mu\right],
\eeq
while the second one does not change.
At small $\mu$, $\mu\ll\sqrt{\xi}/g$ both eigenvalues still cross the
zero at $z=z_0$.

We conclude that in \none theory the critical domain wall
supports four fermion zero modes (at least in some domain of small $\mu$
and large $m$).
 This means that the
\ntwo vector multiplet living on the wall (2 real boson fields + 2 
Majorana fields) is not split
even {\em after} we break \ntwo supersymmetry down to \none. 
The low-energy theory
is still given by (\ref{21theory}). Presumably,  we would feel \ntwo breaking
if we considered higher derivative terms  
on the world volume.
We definitely see   \ntwo breaking
in the spectrum of
  massive excitations   localized on the wall,
with masses of the order of the inverse width of the
wall $\sim 1/R$.

\section{Comments on the literature}
\label{comments}
\setcounter{equation}{0}

The topic under consideration is rather hot;
its aspects have been discussed in the
recent literature in various contexts, including rather exotic, e.g. 
 gauge field localization on  branes
in the framework  of  noncommutative  field theories \cite{R}.
Here we briefly comment on the relation between our results
and those one can fine in the literature.

\subsection{Generalities}
\label{gener}

The most recent revival of the theme of field-theoretic implementation
of D branes and strings can be attributed to
Ref. \cite{DV} which presents an excellent elaboration
of general ideas as to how gauge fields can be localized
on domain walls. A variety of examples are worked out 
providing a clear-cut illustration to the statement \cite{DS}
that localization of the gauge fields requires
confinement in the bulk. It is also explained how this
automatically entails the existence of the flux tubes ending on the walls.
In Sec. 3 of Ref. \cite{DV} the authors construct a model
for a (quasistable) wall-antiwall configuration of a variable thickness
which traps a gauge field in the middle domain.
This model served as an impetus for our construction, which,
being totally different in many aspects, shares a common feature with that
of Ref. \cite{DV} --- the thickness of the
middle domain in our model is a large variable parameter too. 

\subsection{Varying $\Delta m$}
\label{varying}

The proof of the existence of the ``second" modulus
$\sigma (t,x,y)$, dual to the U(1) gauge field on the wall,
 and its massless fermion superpartner
 was based on symmetry arguments
and was independent on the value of the ratio
$\Delta m/\sqrt\xi$. In our model this ratio is large.
However, the U(1) gauge field localization 
must occur for arbitrary $\Delta m/\sqrt\xi$.
This observation perfectly matches the result of Ref. \cite{GPTT},
where a 1/4 BPS solution of 
the string-wall junction type was 
found in a sigma model. In our language the
sigma-model limit corresponds to $\Delta m/\sqrt\xi\to 0$. 
In this limit the photon field and its superpartners
are heavy; upon integrating them out we recover the K\"ahler
sigma model with the Eguchi-Hanson target space for the
remaining light matter fields. It is well-known that
such model  have string-type solitons, 
which can have arbitrary transverse dimension (see Ref.~\cite{AV} for a review
of the so-called semilocal strings).
 
 Turning on $\Delta m \neq 0$ as a small perturbation
produces a small potential on the  target space.
Once the potential is switched on,
domain walls become possible, and one can search for the
string-wall junctions. Strictly
speaking the solution found in Ref.~\cite{GPTT} is somewhat singular,
since   finite-radius strings  exist only in the limit of  massless sigma models
while in this limit there are no domain walls.
In massive sigma models treated
in Ref. \cite{GPTT} the strings are forced
to have vanishing transverse size, and, in fact,  
a ``spike''-type junction was obtained. This is as close as one can get
to \ntwo SQCD string-wall junctions in sigma models.

\subsection{Multiplicity of
the Kaplunovsky-Sonnenschein-Yankielo\-wicz domain walls}
\label{mult-ksy}

In this section we leave the solid ground of weakly coupled models
and venture into uncharted waters of strong coupling.
The issue to be addresses is 
the domain wall
connecting the monopole and dyon vacua in \ntwo SQCD
slightly perturbed by $\mu$Tr$\Phi^2$ term (with SU(2) gauge group).
This problem has been recently studied \cite{KS} by
Kaplunovsky {\em et al.} (KSY for brevity). 
We would like to address the issue of multiplicity of such
domain walls (i.e. the number of distinct domain walls interpolating
between the given vacua,
with degenerate tensions).

First of all, let us summarize what is known about the
multiplicity of the super-Yang-Mills walls at strong coupling.
This question was analyzed by Acharya and Vafa,  Ref.~\cite{AVafa}, 
from the  string theory side. 
Representing the (1+2)-dimensional domain
wall of the super-Yang-Mills theory as a D4-brane wrapped over $S^2$,
the authors found that the U(1) gauge field localized on the wall
is described by  supersymmetric QED
(similar to the construction discussed in Sec. 5.4)
with the Chern-Simons term at level $N$,
\beq
{\cal L}_{1+2} = -\frac{1}{4e^2_{\rm eff}}\, F_{mn} F^{mn}
+\frac{N}{16\pi}\, F_{mn}\, A_k\, \varepsilon^{mnk}+\mbox{ferm. terms}\,,
\label{achvafa}
\eeq
where $N$ is related to SU($N$) of the underlying (1+3)-dimensional
gauge theory.
For the SU(2) gauge group $N=2$. It is well-known that
the level of the Chern-Simons term
determines the number of vacua of the theory --- two in the case 
at hand (the gauge group of the underlying theory is  SU(2)).
The number of vacua in the (1+2)-dimensional effective theory on the wall
{\em is} the number of distinct degenerate domain walls in
(1+3)-dimensional theory (the above  ``two" refers to the 
counting convention explained in Sect. \ref{ntwomultipli}).

Needless to say, this is an index of the underlying theory,
which does not change under continuous deformations of the theory.
Basing on this fact, the domain-wall multiplicity was
calculated directly from field theory
\cite{Ritz:2002fm}, and was found to coincide with the Acharya-Vafa result.

Let us return now to Kaplunovsky {\em et al.}
Since they considered 
 \ntwo SQCD
 perturbed by $\mu$Tr$\Phi^2$ term, and at large $\mu$
this theory smoothly goes into \none gluodynamics,
the index argument tells us that the number of distinct KSY
walls, in the case of the SU(2) gauge group, is {\em two}.
 
Near each vacuum  --- monopole and dyon --- Kaplunovsky {\em et al.} use
  distinct effective low-energy SQED-type
descriptions, e.g. near the monopole vacuum the
low-energy model includes a U(1) gauge  field and its superpartners,
plus  light monopole superfield ${\cal M}\,,\,\, \tilde {\cal M}$. Near the dyon vacuum it is
also a U(1) gauge superfield  (albeit not that of the previous patch)
plus a light dyon superfield ${\cal D}\,,\,\,\tilde {\cal D}$. 
In the intermediate patch the authors keep just one superfield ---
that of $u={\rm Tr}\Phi^2$. It is worth stressing that
no  unified description exists and the consideration 
has to be carried out in three distinct patches separately. 
A {\em unique} solution to the
Bogomolny equations was found.

The KSY solution  bears a remarkable resemblance
to the domain wall in our weakly coupled model. 
Indeed, the KSY wall consists of three domains ---
a broad middle domain of size $\sim \mu^{-1}$, where
the monopole and dyon condensates do vanish,
and two much narrower edge domains (of thickness
$\sim (\mu\Lambda)^{-1/2}$) where the transition from the vacuum value
of the condensate ${\cal M}\tilde {\cal M}$ (or ${\cal D}\tilde {\cal D}$)
to zero occurs. So, why we speak of an unsolved problem?

That's because the KSY solution shows no sign of {\em two}
distinct domain walls. As was explained above,
the double degeneracy of the domain wall is a must.
 Since the index does not depend
on the value of $\mu$, the limit of small $\mu$
considered in Ref.~\cite{KS} must exhibit the same number of domain
walls as the one emerging in the large $\mu$ limit.

What is lacking in   Ref.~\cite{KS}?
An obvious  analogy with  our weakly coupled model
prompts us that the lacking element
 is the analysis of moduli (or quasimoduli)
fields localized on the wall.

For the domain wall at weak coupling, 
considered in the present paper, there is an unambiguous
supersymmetry-based  argument 
proving the double degeneracy.
In the \ntwo  limit our BPS wall belongs to the short representation
of \ntwo superalgebra, i.e. we have two boson + two fermion states (4 fermion
zero modes). When we break \ntwo down to \none our wall is still
\none BPS --- it  belongs to a short representation of
\none superalgebra. This is possible only if we have {\em two} \none
short multiplets  because the number of states cannot
discontinuously change. One of these supermultiplets
is the translational modulus plus its superpartner,
another is the phase field $\sigma $ and its superpartner.

For the KSY wall, generally speaking,  this argument 
does not apply because we
do not  have \ntwo  limit: the \ntwo  breaking parameter (\ref{n2brw})
is never small. 
Why, nevertheless, we suggest that the missed multiplicity
of the KSY walls might be associated with a 
missed (quasi)modulus?

In the KSY problem, there are two distinct {\em dynamically generated}
phase symmetries --- one associated with the phase 
rotations of ${\cal M}$, another with the phase 
rotations of ${\cal D}$. At the same time the
U(1) gauge field is single
(though it is described  differently  in the two edge patches).
There are no massless particles in either of the vacua.
So far, all this is perfectly   parallel to what we have in our model.

Now the two theories divorce.
Ours has an exact global U(1), unbroken in the vacua and spontaneously
broken on the wall, which results  in the strictly massless
modulus $\sigma (t,x,y)$. The Seiberg-Witten theory perturbed by
$\mu$Tr$\Phi^2$ (the KSY case)
has no strictly conserved U(1).
Due to the full similarity in the description of
the edge domains, one may expect, however,
the emergence of a quasimodulus localized on the KSY wall.
We will call it $\tilde \sigma (t,x,y)$. 

To make quasimodulus $\tilde{\sigma}$  massive we have to
include a periodic superpotential, of the type  $\cos (\tilde{\sigma})$
giving rise to a scalar potential of the type
${\sin}^2(\tilde{\sigma})$ which
has two vacua. 
This is quite a general argument
based only on periodicity in $\tilde{\sigma}$
and on the existence of the quasimodulus field localized
 on the KSY domain wall,
 our basic conjecture in this Section.

The quasimodulus field with the target space on $S_1$,
(with a non-vanishing    mass)  can be transformed into
 a dual (2+1)-dimensional
QED with a Chern-Simons term. 
In this way we match with the Acharya-Vafa analysis.

A slightly different argument for this quasimodulus field
$\tilde \sigma (t,x,y)$ is as follows. In each of the edge
domains of the wall (i.e. near the monopole and dyon vacua)
the underlying theory is approximately \ntwo.
The parameter governing the breaking of  \ntwo  is of the order
$\sqrt{\mu/\Lambda}$. In the middle of the
wall, the breaking of \ntwo is stronger,
but it is natural
to think that the phase field $\tilde \sigma$ is 
essentially disassociated
from the middle domain. Then the effective theory  
on the wall must be close to (2+1)-dimensional \ntwo,
which would require two real massless bosonic moduli.
The breaking of \ntwo splits the supermultiplet into two
(2+1)-dimensional \none supermultiplets as
now there is no exact U(1) to prevent the splitting.
A natural estimate for the mass of the quasimodulus
$\tilde\sigma$ is then $$m_{\tilde\sigma}\sim \sqrt{\mu\Lambda}\,
\sqrt{\mu/\Lambda}\sim \mu\,.$$
The dual (2+1)-dimensional SQED
will have the Lagrangian (\ref{achvafa}) with $e^2_{\rm eff} =\mu /\kappa$
where $\kappa$ is a dimensionless constant of order one.
The value of the constant in front of $ F_{mn}F^{mn}$
will be  in accord with our result (\ref{rel21cc}), since the thickness
of the KSY wall is $\sim \mu^{-1}$.

\section{Brief conclusions}

We suggest and work out a  model 
which seems to be a good prototype for
studying basic properties of D brane/string theory
in the field-theoretic setting.
Our model is weakly coupled, fully controllable
theoretically and possesses
both critical walls and strings.

Then, using our model as a tool we addressed
in a fully quantitative manner
the following long-standing issues:

(i)  gauge field localization on the wall;

(ii)  the wall-string junction (i.e. a flux tube coming from infinity
and ending on the all).

We confirm  that   1/2 BPS domain wall   does
localize a U(1) gauge field;  the charge which presents
 the source for this
field is confined in the bulk. 

We find that an 1/2 BPS
flux tube coming from
infinity does indeed end on the above wall.
The wall-string junction is 1/4 BPS. 

A task which remains for the future is 
(the quantitative analysis of) localization of non-Abelian gauge fields
on the wall and related flux-tube--wall junctions.

\section*{Acknowledgments}

We are grateful to Alexander Gorsky, Adam Ritz and 
Arkady Vainshtein for  helpful discussions. 
The work of M.~S. is supported in part by DOE   grant 
DE-FG02-94ER40823, A.~Y.
is   supported in part by the Russian Foundation for Basic
Research   grant No.~02-02-17115, by INTAS grant
No.~00-00334 and by Theoretical Physics Institute
at the University of Minnesota.

\end{document}